\documentclass[fleqn,usenatbib]{mnras}

\usepackage{newtxtext,newtxmath}

\usepackage[T1]{fontenc}

\DeclareRobustCommand{\VAN}[3]{#2}
\let\VANthebibliography\thebibliography
\def\thebibliography{\DeclareRobustCommand{\VAN}[3]{##3}\VANthebibliography}

\usepackage{natbib}

\usepackage{graphicx}	
\usepackage{amsmath}	

\title[Dynamically cold discs in simulated high-z galaxies]{Dynamically cold discs in high-redshift galaxies: comparison between ALMA observations and TNG50}
\author[Y. He et al.]{
Yi He$^{1}$\thanks{E-mail: linn@astro.rug.nl},
Qi Guo$^{2}$,
Filippo Fraternali$^{1}$,
Hang Yang$^{2}$,
Shihong Liao$^{2}$
\\
$^{1}$Kapteyn Astronomical Institute, University of Groningen,Postbus 800, 9700 AV Groningen, The Netherlands\\
$^{2}$National Astronomical Observatories, Chinese Academy of Science, 20A Datun Road, Chaoyang District, Beijing, 100101, China\\
}

\date{Accepted XXX. Received YYY; in original form ZZZ}

\pubyear{\the\year{}}

\begin{document}
\label{firstpage}
\pagerange{\pageref{firstpage}--\pageref{lastpage}}
\maketitle

\begin{abstract}
Observations of highly rotationally supported gas discs in high redshift ($z$ > 3) star-forming galaxies challenge our understanding of galaxy formation, as the prevailing view holds that galaxies in the early universe are dynamically hot due to frequent mergers, gas accretion, and strong stellar feedback. 
We examined the kinematic properties of massive ($M_{\star} \geq 10^{10}\,M_{\odot}$) star-forming galaxies in the TNG50 cosmological hydrodynamical simulation in the redshift range $3\leq z \leq 5$. Mock emission line datacubes were constructed and analysed using the same methodology as for [CII] observations with ALMA.
We measured the ratio of the gas rotation velocity ($V$) to velocity dispersion ($\sigma$) finding that most galaxies have $V/\sigma\sim$ $2-3$, lower than observed. However, a few simulated galaxies show $V/\sigma$ > 5. Such "cold" discs, selected at $z=4$, remain dynamically colder than most of the TNG population across $z=3-5$. A galaxy with $V/\sigma\gtrsim10$ appears in a transient phase that lasts $\leq200$ Myr. 
Dynamically cold disc formation in TNG50 is promoted by gas accretion with angular momentum aligned with the pre-existing disc, while most galaxies undergo misaligned accretion. 
Dynamically cold discs also show lower mass accretion rates and better aligned stellar and dark-matter angular momentum vectors. 
By tracing their evolution to $z = 0$, we find that one-third become massive disc galaxies and two-thirds become ETGs.
\end{abstract}



\section{Introduction}

In the standard scenario of galaxy formation, gas, upon being drawn into the gravitational potential wells of dark-matter halos, collapses and transitions to star formation when it reaches the requisite cool and dense state \citep[e.g.][]{white_core_1978, 1991ApJ...379...52W,springel_cosmological_2003}. 
Galaxies are expected to go through a more turbulent phase at high redshift compared to nearby galaxies because they are fed by cold gas streams with clumpy structures and experiencing powerful outflow from high star formation rates and associated stellar feedback \citep{dekel_cold_2009,hopkins2018fire}. Frequent merger events also play crucial roles in the enhancement of gas velocity dispersion and disc instability \citep{stewart2008merger,hopkins2009disks,forster_schreiber_sinszc-sinf_2014}. 

A straightforward measurement that quantifies the importance of ordered motions, due to rotation, with respect to random motions in a galaxy is the $V/\sigma$ ratio. Initially introduced for the characterisation of the dynamical state of the stellar component in local galaxies \citep[e.g.][]{Scorza&Bender1995, Krajnovic+2008}, it has then been extensively applied to describe the dynamical state of the gas component in high-$z$ galaxies \citep[e.g.][]{ForsterSchreiber+2009, rizzo2024alma}. 
Several studies carried out at cosmic noon and using mostly ionised gas tracers have shown a general decrease of $V/\sigma$ with redshift for the average galaxy population \citep{wisnioski_kmos_2015, ubler_evolution_2019}, pointing to an increasing importance of chaotic motions at high $z$. 
The extrapolation of these cosmic-noon results, coupled with predictions from cosmological hydrodynamical simulations, then indicated that $V/\sigma$ should have reached values of order 1 at $z>3$ \citep{Pillepich+2019} with the implication that "cold" discs, i.e.\ plainly rotation-dominated galaxies should be highly uncommon at $z=4$ and above.

The above expectations were recently shaken by a number of observations carried out with ALMA that revealed the presence of several dynamically cold discs at redshifts $z \approx 4-5$, utilising the fine-structure line of singly ionised carbon at 158 $\mu$m ([CII]) \citep{rizzo_dynamically_2020,neeleman_cold_2020,lelli_massive_2021,fraternali_fast_2021,rizzo2021dynamical,2023MNRAS.521.1045R}.
Throughout this work, we use the term “cold” as shorthand for “dynamically cold.”
The galaxies harbouring these gaseous discs have typical stellar masses of $M_{\star}\gtrsim 10^{10} M_{\odot}$ and very high star formation rates, exceeding $1000 \,M_{\odot}\,{\rm yr}^{-1}$ in some cases \citep{2023MNRAS.521.1045R}.
However, the velocity dispersions measured in [CII] in these systems show typical values of $\sigma \approx 50 \,{\rm km}\,{\rm s}^{-1}$, which, in combination with high rotation velocities, leads to high ($> 5$) $V/\sigma$ values, similar to those found in nearby disc galaxies \citep{van2001kinematics, kurapati2018mass, Bacchini+2020, aditya2021cold}. 
since [CII] traces both atomic and molecular gas \citep{Carilli&Walter2013}, such velocity dispersions indicate significantly higher turbulence in these high-$z$ discs with respect to local galaxies. 
However, this increase appears to be easily explained by the enhanced feedback associated with the high star formation rates of these galaxies \citep{2023MNRAS.521.1045R, rizzo2024alma}.
It is important to note that these measurements have been carried out using a 3D approach to fit emission line data cubes \citep[e.g.][]{teodoro20153d}, a procedure that guarantees a proper treatment of observational biases in determining kinematic parameters. 

The detection of massive disc galaxies at high redshifts poses a challenge to the conventional galaxy formation model.
Although some zoom-in simulations have reported the presence of disc galaxies with high $V/\sigma$ values at high redshifts \citep{kohandel2024dynamically}, there is a tendency for these discs to be short-lived and thus transient phenomena \citep{kretschmer2022origin}, in apparent contradiction with the large fraction of discs seen in the data.
The relatively low number of ALMA observations carried out with sufficient resolutions and signal-to-noise (S/N) ratios to allow for a proper determination of the gas kinematics does not allow a full statistical assessment of the fraction of discs. 
However, a number of considerations indicate that dynamically cold gaseous discs are common in galaxies up to $z \approx 5$.
First, when galaxies are selected from the ALMA archive with the only criterion of data quality, in particular high-spatial resolution, the norm is to find regularly rotating discs \citep{2023MNRAS.521.1045R}.
In fact, high spatial resolution is critical to distinguish discs from interacting systems \citep{Rizzo+2023}.
Second, most galaxies in the $z=4-5$ range are classified either as sub-millimetre galaxies (SMGs, \citealt{fraternali_fast_2021}) or dusty star-forming galaxies (DSFG, \citealt{rizzo2021dynamical}) or active galactic nuclei (AGNs, \citealt{lelli_massive_2021}). High star formation, potential post-merger status, and AGN feedback are all phenomena that should bias towards a higher dispersion with respect to the bulk of the galaxy population.
Third, despite the high resolution of the ALMA data and the 3D approaches used, unresolved streaming motions and residual beam smearing effects also tend to increase the estimated velocity dispersion. 
It is therefore unlikely that typical galaxies at $z=4-5$, at least in the stellar mass regime of $M_{\star}\gtrsim 10^{10} M_{\odot}$, will have significantly lower values of $V/\sigma$ than measured so far, i.e.\ gas discs in massive galaxies are common at these redshifts.
Therefore, it is of paramount importance to establish whether dynamically cold discs are formed in current cosmological hydrodynamical simulations at high redshift and, if so, how common they are.

To address these questions, in this paper, we use the cosmological hydrodynamical simulation TNG50-1 \citep{weinberger2016simulating,pillepich2018simulating,2019MNRAS.490.3234N,Pillepich+2019}, which provides high resolution and statistical capabilities to investigate how common high-redshift cold discs are and what mechanisms drive their formation.
Previous work with TNG50 concentrated on the kinematic properties of the average population \citep{Pillepich+2019}, without investigating the presence of possible outliers at high $V/\sigma$.
Moreover, the comparison was made using observations of the ionised (warm) gas component, which typically exhibits higher velocity dispersion than the cold gas \citep[see e.g.][]{2021ApJ...909...12G,2022MNRAS.514..480E,kohandel2024dynamically,rizzo2024alma}. 
Mock data were also produced to mimic KMOS H$\alpha$ observations \citep{2019MNRAS.490.3234N, Pillepich+2019}, which have lower S/N and resolution than ALMA [CII] data.
Here, instead, we generate mock emission-line datacubes from simulated galaxies to mimic the ALMA [CII] observations at $z=3-5$ and carry out a one-to-one comparison with observed galaxies.
Additionally, we analyse the evolution of the accretion of gas, stars, and dark matter in TNG50 galaxies to uncover the physical processes driving cold disc formation. 

The structure of this paper is as follows: in Section~\ref{section:method}, we describe sample selection, kinematic analysis, and mock observations. 
The kinematic results and the evolutionary dynamics of different components are presented in Section~\ref{section:result}. In Section~\ref{section:discussion}, we compare our findings with previous studies, explore the formation pathways of dynamically cold discs, and discuss their potential fate. We summarise and conclude in Section~\ref{section:conclusions}.


\section{Method}
\label{section:method}
\subsection{TNG50 simulations and halo finder}

The IllustrisTNG project \citep{2018MNRAS.480.5113M,2018MNRAS.477.1206N,2018MNRAS.475..624N,2018MNRAS.475..648P,2018MNRAS.475..676S} is a suite of cosmological magnetohydrodynamical simulations performed using the moving-mesh code {\sc AREPO} \citep{2010MNRAS.401..791S}. It consists of three distinct volumes: TNG50, TNG100, and TNG300 with side lengths of approximately 50, 100, 300 comoving Mpc, respectively. The adopted cosmological parameters are consistent with the Planck15 results \citep{2016A&A...594A..13P}: $\Omega_{m,0} = 0.3089$, $\Omega_{b,0} = 0.0486$, $\Omega_{\Lambda,0} = 0.6911$, $\sigma_{8}=0.8159$, $n_{s}=0.9667$, $h=0.6774$. In this paper, we use the highest-resolution simulation of TNG50, TNG50-1, which is also the most computationally demanding \citep{2019MNRAS.490.3234N,Pillepich+2019}. TNG50-1 contains $2\times2160^3$ resolution elements, providing details on the structures of galaxies and the dynamics of gas, stars, and dark matter. This simulation has a resolution in mass for baryons and dark matter, respectively, of $m_{\rm baryon} = 5.7\times10^{4}\,M_{\odot}/h$, $m_{\rm DM}=3.1\times10^5\, M_{\odot}/h$.

Halos are identified using the standard Friends-of-Friends (FoF) algorithm \citep{1985ApJ...292..371D}, with a linking length parameter of $b = 0.2$, expressed in units of the mean interparticle distance. 
In the FoF algorithm, particles separated by less than this distance are linked together, and all connected particles form a group that is identified as a dark matter halo. 
Gravitationally bound substructures (i.e. subhalos) are then identified within the FoF groups using the SUBFIND algorithm \citep{2001MNRAS.328..726S,2009MNRAS.399..497D}. Galaxies are defined as subhalos that also contain stellar particles. 
The stellar mass of galaxies is determined by the total mass of stellar particles within the subhalo. Galaxies residing in the deepest gravitational potential wells of their respective groups are classified as central galaxies, while others are referred to as satellites.

In this work, we use the trees created by SubLink \citep{rodriguez-gomez_merger_2015} to identify progenitors and descendants, allowing us to study the evolution of the halos and the galaxies embedded within them. SubLink constructs merger trees at the subhalo level by identifying descendant candidates based on common particles in subsequent snapshots, and assigns a unique descendant using a merit function that considers the binding energy rank of each particle.

\subsection{Criteria for sample selection from simulations}

Our main goal here is to compare simulated galaxies with observed massive star-forming galaxies that exhibit a cold gas disc at high redshifts. 
Thus, we concentrate on central galaxies with stellar masses between $10^{10}\ M_{\odot}$ and $10^{11}\ M_{\odot}$ at redshifts ranging from 3 to 5. Such a redshift range includes nine snapshots of the TNG simulation. 
Following \citet{Pillepich+2019} and \citet{hemler2021gas}, we set a threshold to distinguish star-forming galaxies from non-star-forming ones. This threshold is defined as 0.5 dex below the median logarithm of the specific star formation rate (sSFR) of the entire population in the simulation at a given stellar mass. Galaxies with sSFR values above this threshold are classified as star-forming. In this initial selection, we identify star-forming galaxies at each snapshot based on their instantaneous star formation rate (SFR) and stellar mass. After the application of these selection criteria, our dataset comprises a total number of 451 galaxies, which is the sum of galaxies from all snapshots. See Table~\ref{table:table1} for the detailed number of galaxies in each snapshot. It is important to note that not all selected galaxies remain star-forming over the entire redshift range. The cold gas masses of our sample range from $10^{10}\ M_{\odot}$ to $10^{11.5}\ M_{\odot}$, which is comparable to the gas masses reported in \citet{Roman-Oliveira+2024}.

\begin{table*}
\centering
\begin{tabular}{|c|c|c|c|c|c|c|c|c|c|}
\hline
{Snapshot number} & 17 & 18 & 19 & 20 & 21 & 22 & 23 & 24 & 25 \\
\hline
{Central redshift} & 5.00 & 4.66 & 4.43 & 4.18 & 4.01 & 3.71 & 3.49 & 3.28 & 3.01 \\
\hline
{Number of galaxies} & 15 & 17 & 28 & 37 & 39 & 53 & 68 & 83 & 111 \\
\hline
\end{tabular}
\caption{Number of star-forming galaxies in each snapshot and corresponding redshift based on our selection criteria. The snapshot numbering follows the default scheme of the TNG50 simulation.}
\label{table:table1}
\end{table*}

\begin{figure*}
\centering
\includegraphics[width=\textwidth]{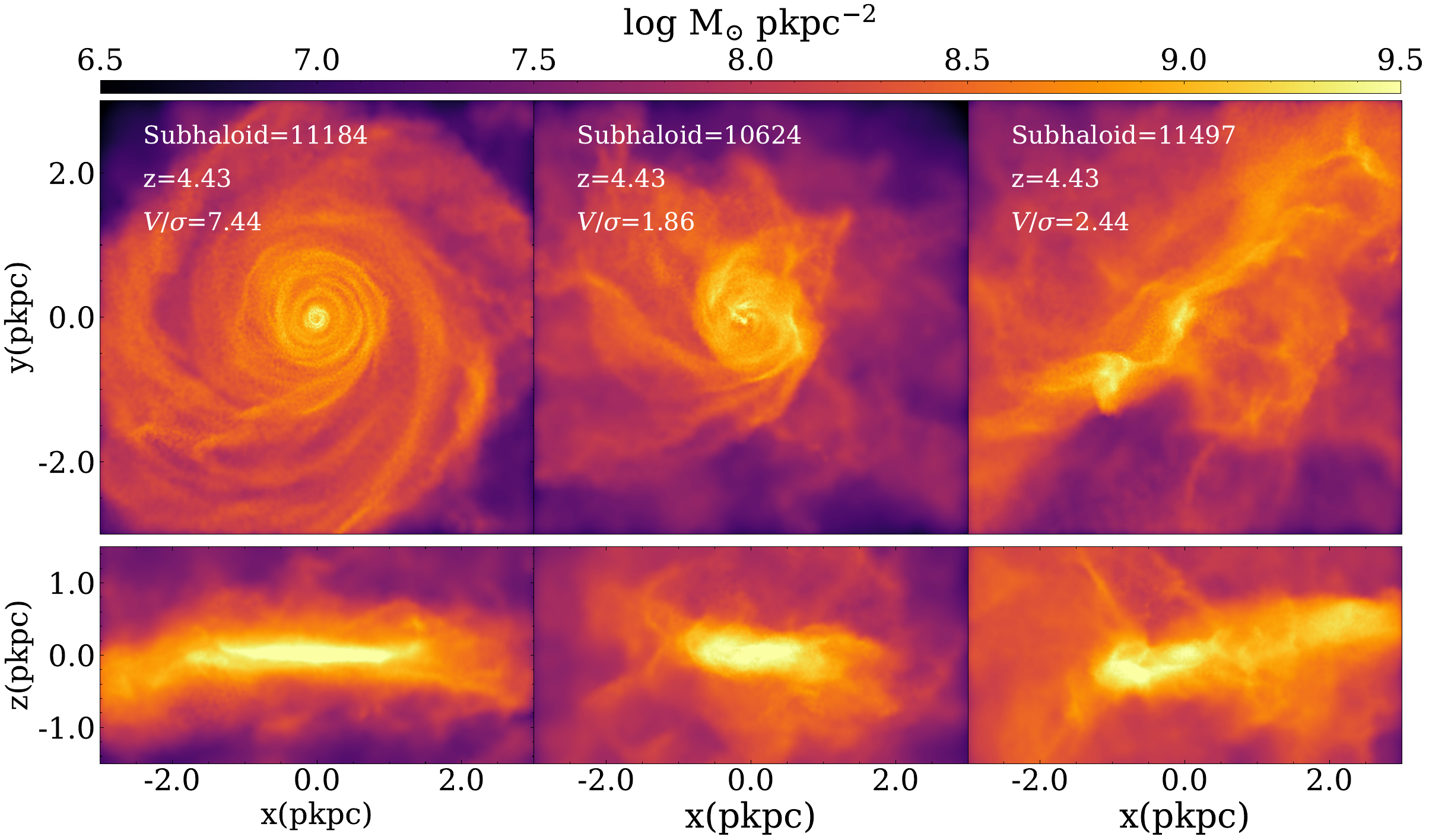} 
\caption{Gas surface densities of neutral hydrogen in three simulated galaxies at $z=4.43$, from TNG50. This figure is rendered
using the Py-SPHViewer code \citep{benitez-llambay_2015}. The leftmost panel shows a cold disc with high $V/\sigma$, while the middle and right panels show two chaotic galaxies, with a $V/\sigma$ close to the average value in TNG50 massive galaxies at this redshift. The upper and lower panels show the face-on and edge-on views, respectively. }\label{fig:fig1}
\end{figure*}

\subsection{Gas kinematics from simulations}
\label{subsection:Kinematic_definition}
To compare observations of cold gas in galaxies at $z \approx 4$, mostly in [CII], with simulated galaxies, we need to extract a comparable gas component from the TNG50 simulation. [CII] serves as a tracer of cold gas, which means both neutral atomic gas and molecular gas \citep{2015MNRAS.449.2883G,tarantino2021characterizing,ramos2021diagnosing,2025A&A...693A.119C}. However, due to the cooling model in TNG50, gas cells do not cool below $10^4$ K, roughly the temperature of the warm neutral medium, making it impossible to directly extract [CII] emission. 
Instead, our strategy has been to estimate the density of neutral hydrogen in atomic and molecular forms ($\rm{HI+H_{2}}$), as [CII] traces both components, making it unnecessary to distinguish between them. 
Fortunately, there are public codes available to estimate the abundance of atomic and molecular hydrogen in large-scale cosmological simulations \citep{2019MNRAS.487.1529D,stevens2019atomic,2019MNRAS.482..821O}. 
These packages estimate the partitioning of gas into atomic and molecular hydrogen using sub-grid prescriptions based on local gas properties such as density, metallicity, ultraviolet radiation field, and/or pressure, calibrated against observations and high-resolution simulations. 
Despite the distinction between HI and H$_2$ can depend significantly on these assumptions, the sum of the two components, which we use here, is robust \citep{Marasco+2025}.
For our analysis, we model the neutral hydrogen gas through the Hdecompose package, which employs the methods of \cite{2013MNRAS.430.2427R} to calculate the neutral hydrogen fractions. 

From the gas cells in our simulated galaxies, we extract two fundamental properties: rotation velocity ($V$) and velocity dispersion ($\sigma$), which we then use to calculate the $V/\sigma$ ratio, in line with observational approaches. 
This ratio is an effective metric for quantifying the level of ordered motions (rotation) with respect to disordered ones. 
Although this method works well for galaxies in equilibrium, not all galaxies in our sample are in such a state. 
Therefore, in addition to $V/\sigma$, we use a 3D asymmetry parameter (Section~\ref{subsection:Asymmetric}), based on observational properties, to provide a comprehensive analysis of the gas distribution and kinematics and ensure a robust comparison with observational data. 

We calculate $V/\sigma$ using gas located within a cylinder perpendicular to the galactic disc plane, with a radius of 5 pkpc and a height of 4 pkpc. These sizes ensure that most of the cold gas associated with the galaxy discs is included.
The orientation of the disc plane is determined by calculating the angular momentum direction of all gas cells, within a 10 ckpc radius. We derive rotation velocities at different radii determined from the tangential velocity of gas cells in bins 0.7 pkpc wide. This method is different from that used by \cite{Pillepich+2019}, which derived the rotation velocity from edge-on projections. In real observations, the derivation of rotation curves of edge-on galaxies requires complex analysis techniques, such as envelope tracing \citep{1979A&A....74...73S,2004MNRAS.352..787K,2011A&A...531A..64F} to overcome projection effects. Deriving rotation velocities directly from an edge-on projection without applying such correction is not recommended, and it can lead to a significant underestimation of the rotation velocity, especially in the inner disc \citep{2006AJ....132.1426S,2010A&A...515A..62O}. Our estimate is instead closer to the observational technique of the tilted-ring model where rotation is determined by averaging the gas velocity in concentric elliptical annuli and correcting for the inclination of the annulus along the line of sight \citep{begeman1989hi,de2008high}.

Various definitions of rotation velocity are used in observational studies when calculating the global $V/\sigma$, in particular the maximum value of the rotation curve $V_{\mathrm{max}}$, the external rotation velocity $V_{\mathrm{ext}}$, the value corresponding to the flat part of the rotation curve $V_{\mathrm{flat}}$ \citep{neeleman_cold_2020,lelli_massive_2021,rizzo_dynamically_2020,fraternali_fast_2021,rizzo2021dynamical,2023MNRAS.521.1045R}. Here, we define the rotation velocity as the average value between $V_{\mathrm{max}}$ and $V_{\mathrm{flat}}$. To estimate the latter, we calculate the differences between consecutive pairs in the data and select the data point corresponding to the smallest difference as $V_{\mathrm{flat}}$. Our choice for $V$ will select $V_{\mathrm{flat}}$ for galaxies with rising rotation curves, while a somewhat larger velocity in galaxies with an inner declining curve. 

We calculate the velocity dispersion of the gas in each radial bin as the standard deviation of the velocity component perpendicular to the disc. We use the same radial bins as the rotation curve, similar to the second moment of the velocity distribution calculated from the tilted-ring model in observations (in high spatial resolution data where the beam smearing effect can be neglected). 
The global velocity dispersion ($\sigma$) of the galaxy is the mean value over the disc. 
We also tested the extraction of velocity dispersion from the radial velocity distribution of the particles in our simulated galaxies, instead of the vertical distribution, and found similar values. We note that both rotation velocity and velocity dispersions are weighted by the neutral hydrogen density of the particles in each radial bin. Our choice of radial bin size is based on the typical spatial resolution of ALMA high-redshift [C II] observations ranging from $\sim0.1''$ to $0.24''$, corresponding to $\sim0.6$–1.5 kpc at $z \sim 4$ – 5 \citep{rizzo_dynamically_2020,neeleman_cold_2020,lelli_massive_2021,fraternali_fast_2021,rizzo2021dynamical,2023MNRAS.521.1045R}.

The error in the rotation speed is determined by calculating the standard deviation within each radial bin, i.e. the same bins used for constructing the rotation curve, while the uncertainty of global $V$ is defined as the mean error between $V_{\mathrm{max}}$ and $V_{\mathrm{flat}}$. 
The error in the global $\sigma$ is defined by the standard deviation of the velocity dispersion values across the disc.
 
Using the above procedure, we calculated the values of $V/\sigma$ for our entire sample of simulated galaxies. 
We then extracted the median and standard deviation of $V/\sigma$ in the nine snapshots. Galaxies with $V/\sigma$ values greater than one standard deviation above this median are classified as galaxies with dynamically cold discs. As an example, Fig.~\ref{fig:fig1} shows two projections of the gas surface densities of three simulated galaxies using the Py-SPHViewer code \citep{benitez-llambay_2015}. 
The three galaxies belong to different categories: a cold disc on the left and two more normal TNG50 galaxies, all at $z=4.43$. In the face-on view, the galaxy with a cold disc has a clear spiral structure. Its spiral arms, like those of other simulated cold discs, appear tightly wound in the central part, suggesting that it might be in a transient phase \citep{2011MNRAS.410.1637S,2012MNRAS.421.1529G}. In the edge-on view, the galaxy with a dynamically cold disc exhibits a geometric thin disc that extends over several kiloparsecs. In contrast, the other two star-forming galaxies, selected to have a $V/\sigma$ at this snapshot close to the population average, are more chaotic, hotter, and less structured. The galaxy in the middle panel appears compact and surrounded by an unsettled envelope, while the galaxy in the right panel shows multiple clumps of cold gas, pointing at a possible interacting system. We discuss this further in Section~\ref{subsection:discform}.

\subsection{Mock datacubes of simulated galaxies}

With the aim to achieve a reliable comparison between simulated and observed galaxies, we have created mock observations of our TNG50 galaxies that are as much as possible similar to actual observations. 
In this study, our aim was to mimic the properties of the ALMA [CII] observations of galaxies at $z \approx 4.5$, in particular those presented in \cite{2023MNRAS.521.1045R}.
Thus, we have generated mock neutral hydrogen datacubes that contain both atomic and molecular gas. 
The creation of such mock datacubes allows us to apply the same analysis methods used on actual observations to the simulations and, at the same time, to test the accuracy of the observational results.
In particular, we compared the $V/\sigma$ values extracted directly from simulations with those derived from mock observations, helping to verify the accuracy of the fitting methods employed in real observations (Section~\ref{subsection:fitting method}). 
Additionally, we used our mock data to examine whether the gas distribution and kinematics in the simulated galaxies display a level of (a)symmetry comparable to that observed in real galaxies (Section~\ref{subsection:Asymmetric}). 

\begin{figure*}
\centering
\includegraphics[width=\textwidth]{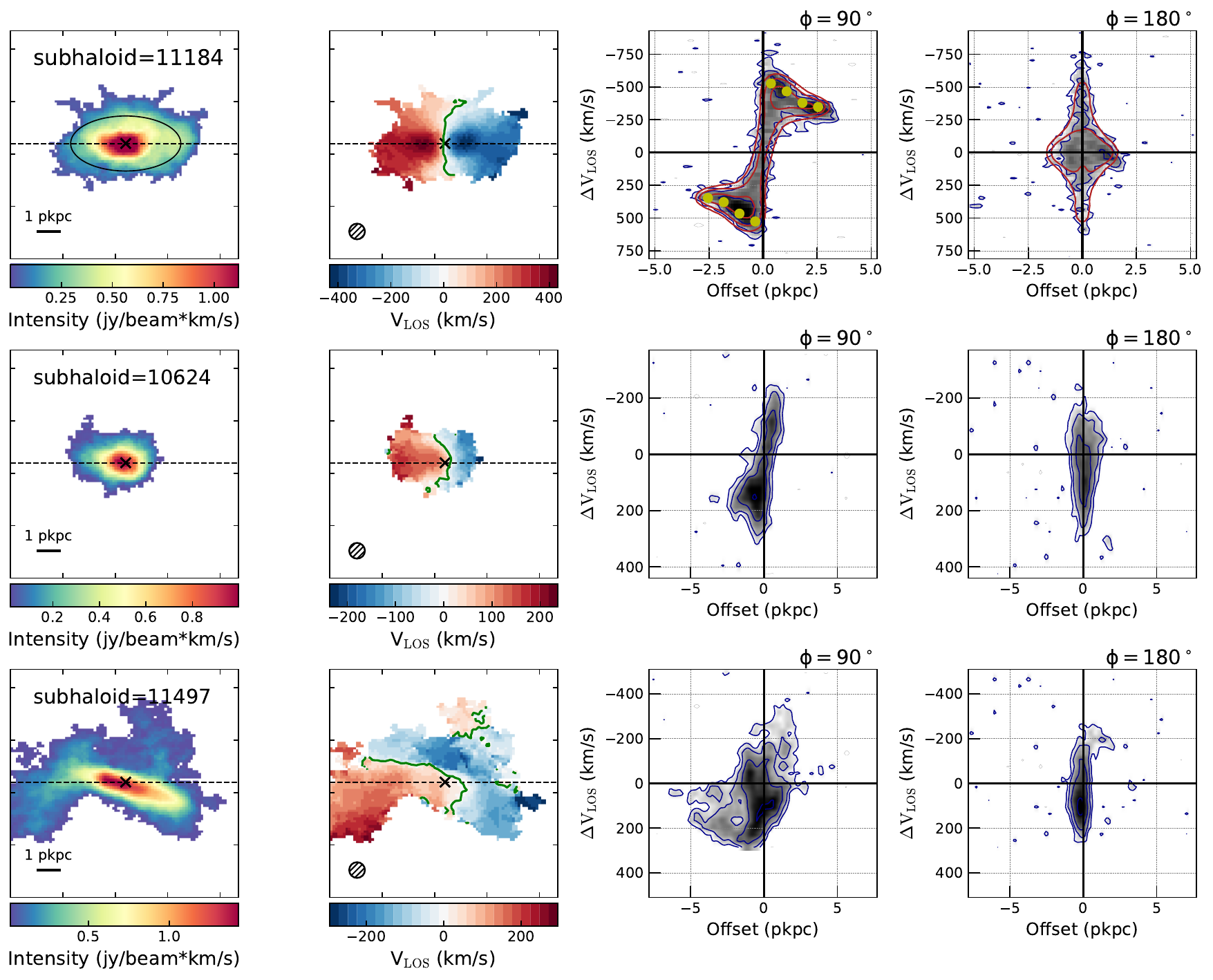} 

\caption{Intensity map, velocity field and position-velocity slices (along the major, $\phi=90^{\circ}$, and minor, $\phi=180^{\circ}$, axes), extracted from our neutral hydrogen mock observations for the three selected TNG50 galaxies presented in Fig.~\ref{fig:fig1} on the $z=4.43$ snaphot. 
The top row displays the cold disc galaxy, while the second and third rows show two representative chaotic galaxies. The physical scale bar and beam are included in the bottom left of intensity maps and velocity field panels, respectively.}
\label{fig:fig2}
\end{figure*}

For each simulated galaxy, we use the Martini modular PYTHON package \citep{2019MNRAS.482..821O} to generate mock emission line observations. This process employs the Hdecompose function to calculate the neutral hydrogen fraction in every cell. 
Although Martini is typically used to produce HI datacubes, i.e.\ containing 21-cm emission from atomic hydrogen, we extend its application here. 
Our goal is to create synthetic data that mimic the [CII] emission, which traces both atomic and molecular cold gas.
To achieve this, we use the total neutral hydrogen mass instead of HI mass to generate the mock datacubes, under the assumption that the intensity of neutral hydrogen emission is proportional to the neutral gas surface density, in the same way that HI intensity is proportional to the HI surface density.
The resulting mock spectroscopic datacubes contain, as customary, two spatial axes and one spectral axis. We put all our galaxies at the arbitrary distance of 5 Mpc, with a pixel size of 5 arcsec. Each datacube is then convolved with a 30-arcsec circular Gaussian function to mimic the observational beam, reaching a physical size of 720 pc, which makes them consistent with the resolution of well-resolved galaxies observed with ALMA at $z = 4-5$ \citep{lelli_massive_2021,2023MNRAS.521.1045R}.
The velocity channels in the mock datacubes are set to a width of 30 km/s, which is the typical ALMA channel width for these types of observations, and the cube contains 64 channels, sufficient to encompass the full velocity range of each galaxy. 
We set all mock observations at the average inclination angle of a random distribution of inclined discs, $i = 60^{\circ}$, and at a position angle $\mathrm{\phi} = 90^{\circ}$. 

To simulate realistic observational conditions typical of interferometric data, we generated a noise cube based on an HI datacube (NGC 5055), extracted from the HALOGAS survey \citep{2011A&A...526A.118H}. 
We expand the noise data by repeating this datacube until it covers the full spectral extent of the mock datacube. 
After this extension, we trim the noise cube to precisely match the spatial dimensions of the mock datacube. 
We chose to use noise from real HI datacubes as it reproduces the spatially correlated noise structure characteristic of CLEANed interferometric spectral-line observations, like the ALMA [CII] data \citep[e.g.][]{Killi+2024}.
The main reason to prefer HI over [CII] data is the larger spatial and spectral coverage of the former, which minimises pattern repetition, providing a more realistic noise distribution than random noise.
To ensure consistent resolution between the noise and mock datacubes, we apply a smoothing process to the noise cube, adjusting it to match the beam width of the mock datacube.
The mock datacube is then added to this noise cube, and the mock line emission is rescaled to achieve an average S/N per channel of 5. 
We calculate the S/N as the sum of the intensity inside a mask built using $^{\rm 3D}$Barolo divided by the standard deviation of the smoothed noise cube. 
As above, we use "SEARCH" in $\mathrm{^{3D}Barolo}$, with parameters SNRCUT=3 and GROWTHCUT=2, to construct the masks.
The value of $\rm{S/N} = 5$ is chosen to mimic the typical S/N of well resolved ALMA data at $z = 4 - 5$ \citep{lelli_massive_2021,2023MNRAS.521.1045R}.

A visualisation of the final neutral-hydrogen mock datacubes is given in Fig.~\ref{fig:fig2}, using the same three galaxies shown in Fig.~\ref{fig:fig1}. 
The two leftmost columns display the intensity map and velocity field of the galaxies, while the third and fourth columns show two position-velocity diagrams extracted from the datacube along the major axis and the minor axis, respectively. The top row showcases the cold disc galaxy, left panel in Fig.~\ref{fig:fig1}, while the mid and bottom rows feature the other two more chaotic systems, shown, respectively, in the central and right panels in Fig.~\ref{fig:fig1}. The position-velocity diagram along the major axis of the top row galaxy displays the typical shape of a disc in differential rotation. For chaotic galaxies, the position-velocity diagrams show more irregular shapes, indicating complex dynamics.

\subsection{Kinematic modelling of mock datacubes} \label{subsection:fitting method}
To extract the kinematic parameters from our mock observations, we use the 3D tilted-ring modelling tool $\mathrm{^{3D}Barolo}$ \citep{teodoro20153d} . $\mathrm{^{3D}Barolo}$ is a software designed to derive the geometry and kinematics of any disc-like object from emission line datacubes by decomposing the object into concentric rings where the gas moves in circular orbits and is characterised by an isotropic velocity dispersion. "3D" indicates that the software fits the 3D tilted-ring model using the entire datacube \citep[see also other codes in][]{2013MNRAS.429..534D, bouche2015galpak3d}, rather than the standard 2D approach applied to velocity fields. Because it uses all the information present in the data, it performs significantly better than the classical 2D approach when applied to low-resolution observations \citep{2022A&A...667A...5R}. 

We used $\mathrm{^{3D}Barolo}$ to analyse our mock observations similarly to what was done for the ALMA observations. We decomposed galaxies into rings having widths of one beam, which ensures that consecutive rings contain nearly independent emissions. The number of rings is determined by the size of the galaxies and is typically between 3 and 5, again in agreement with the $z=4$ ALMA data \citep{2023MNRAS.521.1045R}. 
We fix the inclination and position angle to the correct values. We fixed the thickness parameter to zero, effectively modelling all galaxies as thin disks. The centre of the rings is set to match the centre of the datacube. 
In all our runs of $\mathrm{^{3D}Barolo}$ we apply an azimuthal normalisation (AZIM for the NORM parameter). 
For the construction of the masks, we adopt the same parameter settings as those used in $\mathrm{^{3D}Barolo}$ to achieve the target S/N level.
At the end, for each ring, we fit rotation velocity and velocity dispersion. This procedure generally produces good fits to all our dynamically cold mock galaxies. 
An example is shown in Fig.~\ref{fig:fig2}, where we overlay our best-fit kinematic model for the cold disc galaxy in red contours in the position-velocity diagrams (top right panels). 
The model contours generally agree with the data contours and reproduce most of the emission. 
The rotation velocity displays a steep peak at the centre, followed by a decline to a nearly constant value in the outer regions.
Note that, due to the limited resolution of the mock data, we do not attempt to model potential warps by varying position angle and inclination between the various rings.
If a warp is present in real and simulated data, it would typically only affect the outer ring, and the global values of $V$ and $\sigma$ would change little; see, e.g., the case of SGP38326-1 in \citet[][]{Roman-Oliveira+2024}.

The error of global $V$ is defined as the quadrature sum of the highest measurement uncertainty at the points of $V_{\mathrm{max}}$ and $V_{\mathrm{flat}}$, and half the difference between these two velocities. 
The error of global $\sigma$ is instead defined as the quadrature sum of the mean value of the measurement error produced by $\mathrm{^{3D}Barolo}$ and the standard deviation of the measured dispersion across the disc. 
The difference in error definitions between simulation and mock observation arises because the simulation data represent an idealised scenario with negligible measurement uncertainties, allowing the errors to be purely statistical. In contrast, mock observations explicitly incorporate observational uncertainties and instrumental effects, thus requiring a combined approach to accurately reflect realistic measurement conditions.
 

\subsection{Asymmetry parameter}
\label{subsection:Asymmetric}

Classifying whether a galaxy has a regularly rotating disc or not cannot be done using $V/\sigma$ alone. 
This parameter gives a quantification of the amount of ordered over random motions only when $\sigma$ can be considered a good tracer of the gas velocity dispersion and there is a clear rotational pattern.
In non-symmetric systems or systems with more than one component, because of the presence of accreting clouds or merger events, the average velocity dispersion of the gas cells will be strongly affected by these phenomena and generally increase, making the $V/\sigma$ ratio decrease even if an underlying rotational supported disc is present. 
Conversely, a chaotic system can also have gradients in velocity that can be misinterpreted as a rotational signal once an azimuthal average is taken.
Overall, the $V/\sigma$ ratio is hardly able to fully capture the complexity of the galaxy dynamics at these high redshifts.

Taking advantage of the fact that a genuine rotating disc will also be symmetric and regular, both in distribution and kinematics, we have decided to pair the $V/\sigma$ classification with a more observational criterion based on a 3D asymmetry parameter.
We computed this asymmetry parameter using the entire datacube, similarly to \cite{deg2023measuring}. 
In their work, they found that, compared with utilising the integral information such as the 2D intensity map or velocity field, the 3D asymmetry parameter works better to classify the regularity of the gas distribution and kinematics, especially for observations with low spatial resolution (less than 5 elements per side). 
Our asymmetry parameter is defined as
\begin{equation}
A=\frac{\sum |I_{i, j, k}-I_{-i, -j, -k}|}{\sum|I_{i, j, k}+I_{-i, -j, -k}|},
\end{equation}
where $(i,j)$ are the spatial pixel indices relative to the centre of the galaxy, $k$ is the spectral channel index relative to the central channel of the galaxy (closer to the systemic velocity), and $I_{i,j,k}$ is the flux of the voxel $(i,j)$ in channel $k$. 
We calculate the gaps between the systemic velocity and the two nearest channels. If one of these gaps is less than one third of the channel width, we designate that channel as the central channel. If both gaps exceed one-third of the channel width, we define both channels as central channels.
The calculation of the asymmetry parameter is done inside a mask, which is constructed using $\mathrm{^{3D}Barolo}$. We use the same approach and parameter as done for real galaxies in \cite{2023MNRAS.521.1045R}, i.e.\ we use the SEARCH option in $\mathrm{^{3D}Barolo}$ with SNRCUT = 3 and GROWTHCUT = 2.5. 
To ensure that the masks are symmetric between the approaching and receding sides of both real and simulated galaxies, we check each pair of corresponding pixels and modify the original mask by masking both pixels in a pair whenever one of them is masked by $\mathrm{^{3D}Barolo}$.
The use of a symmetric mask is crucial, as an asymmetric mask itself will affect the value of the measured asymmetry \citep{deg2023measuring}. 

In Section \ref{sec:Vsigma&asym}, we compare the values obtained from the asymmetry parameter A in simulated and real galaxies and establish a threshold of $A=0.35$ to distinguish between symmetric and non-symmetric systems.


\section{Result}
\label{section:result}
\subsection{$\mathrm{^{3D}Barolo}$ performance in retrieving the correct $V/\sigma$ ratios} 

\begin{figure}
\includegraphics[width=0.5\textwidth]{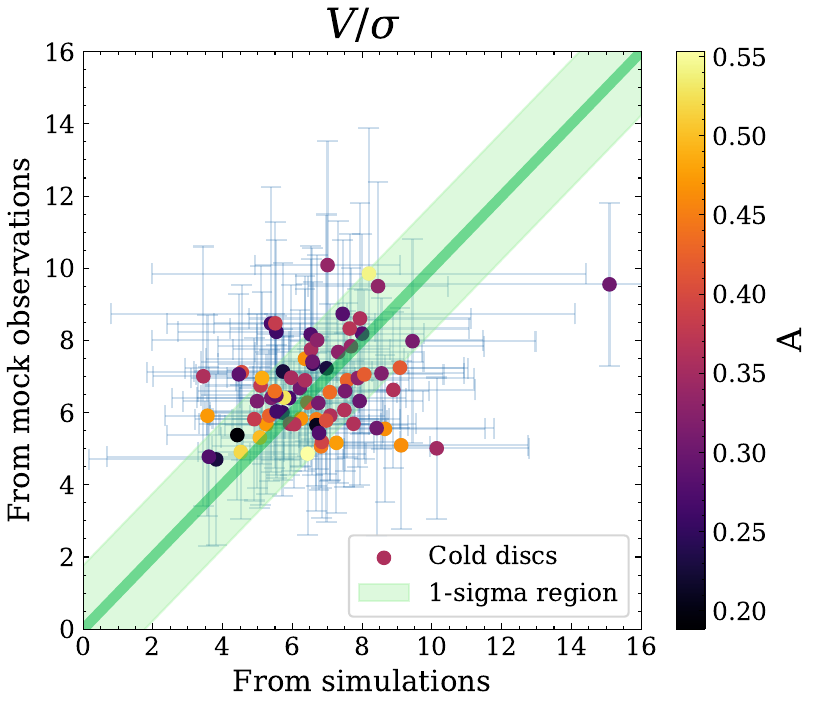}
\caption{Comparison of $V/\sigma$ values of dynamically cold discs in TNG50 obtained either directly from the simulations (x-axis) or from the mock datacubes (y-axis). Each point represents one simulated galaxy, coloured by its asymmetry ($A$) value. Horizontal and vertical error bars indicate simulation and mock uncertainties, respectively. The solid green line marks the 1:1 relation and the dashed line the 1-sigma scatter.}
\label{fig:voversigma_mockvssim}
\end{figure}

Here we show that the fitting technique used for most of the observed ALMA data at $z$ = 4 - 5 to date, $\mathrm{^{3D}Barolo}$, applied to our mock galaxy sample returns correct values for the $V/\sigma$ ratios.
Fig.~\ref{fig:voversigma_mockvssim} compares the $V/\sigma$ values of dynamically cold disc galaxies derived directly from the simulations versus the values obtained by applying $\mathrm{^{3D}Barolo}$ to our mock observations. To further evaluate the accuracy of the kinematic recovery, we also present separate comparisons of $V$ and $\sigma$ in Appendix~\ref{appendix:apenA}.
Clearly, all galaxies scatter around the 1-to-1 line. 
The 1-sigma scatter, defined as the standard deviation of the orthogonal distances of the points from $y=x$, is 1.3 for the full sample. The mean offset is 0.04 in total sample. This relatively narrow band indicates that, overall, $\mathrm{^{3D}Barolo}$ accurately recovers the dynamical properties of most cold disc galaxies. 
Most importantly, there is no systematic shift from the identity line.

As we mentioned in Section~\ref{subsection:Asymmetric}, some galaxies with high $V/\sigma$ exhibit asymmetric structures. 
These galaxies may pose challenges to derive a meaningful $V/\sigma$ ratio from their mock datacubes, given that codes like $\mathrm{^{3D}Barolo}$ rely on the axisymmetry of the disc, which is also assumed to be thin. 
In Fig.~\ref{fig:voversigma_mockvssim}, we show the asymmetry of the galaxies with cold disc by the colour. 
It is clear that $\mathrm{^{3D}Barolo}$ performs satisfactorily in both asymmetric and symmetric cold disk galaxies. 
The mean offset and orthogonal scatter of the subsample with $A$ < 0.35 (symmetric objects) are 0.25 and 1.3. And the offset is mainly caused by the one outlier.

We have investigated some of the outliers in Fig.~\ref{fig:voversigma_mockvssim} to understand what can cause an over- or under-estimation of $V/\sigma$ from the analysis of emission line datacubes. 
The most obvious outlier is the galaxy with $V/\sigma \approx 15$ from the simulation (subhaloid = 36491 at $z = 3.49$). 
This system has a very low velocity dispersion of the cold gas ($\sigma \simeq 23$ km/s). In this case, it is probably the low velocity resolution (30 km/s) of our mock datacube that makes $\mathrm{^{3D}Barolo}$ overestimate $\sigma$ and underestimate the ratio. 
We also examined galaxies that have higher $V/\sigma$ from the mock observations but low actual values in the simulations. 
In these cases, it appears that distortions in the major and minor axes, caused by warp structures or radial flows, make it difficult for $\mathrm{^{3D}Barolo}$ to measure the true kinematic properties.

Overall, we can conclude that the values of the $V/\sigma$ ratios derived from $\mathrm{^{3D}Barolo}$ from real or mock datacubes are compatible with the $V/\sigma$ ratios that we estimate directly from the simulations. 
We can therefore proceed to compare the $V/\sigma$ values that we derived for our TNG50 sample with the literature results from real galaxies.

\begin{figure}
\includegraphics[width=0.48\textwidth]{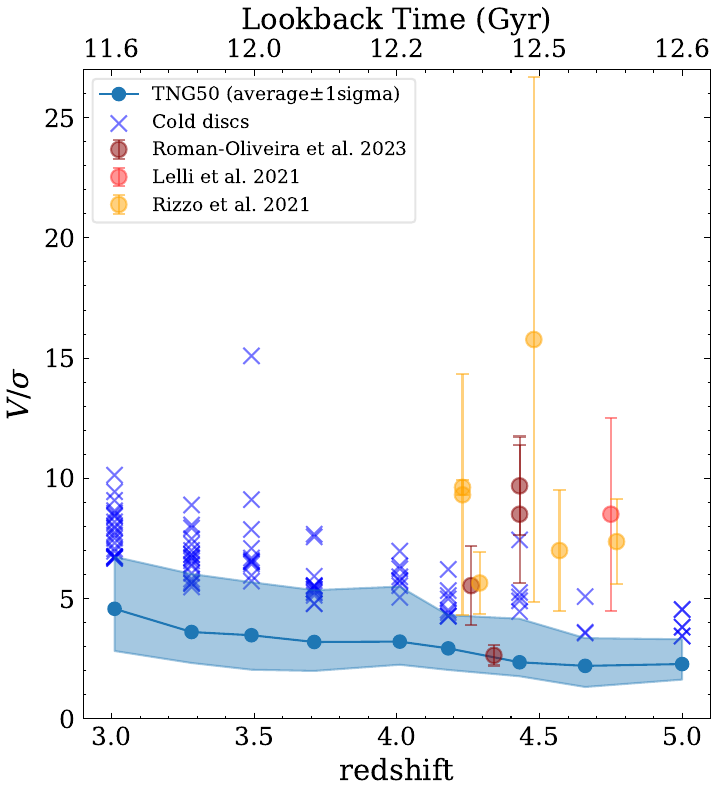}
\caption{The V/$\sigma$ ratio as a function of redshift for cold gas in star-forming galaxies from TNG50, compared to the V/$\sigma$ ratios in real galaxies observed with ALMA in the [CII] emission line. 
The dark blue curve shows the median V/$\sigma$ in all the simulated galaxies in this mass range at the different times of the available snapshots.
The band shows 1-sigma deviation from the median. 
Galaxies with V/$\sigma$ values greater than 1-sigma above the median value are defined as cold discs (blue crosses). }

\label{fig:voversigma}
\end{figure}

\subsection{V/$\sigma$ of the simulated galaxies}\label{sec:Vsigma&asym}

The main findings regarding the $V/\sigma$ classification of TNG50 galaxies are illustrated in Fig.~\ref{fig:voversigma}. 
Here we show the median values of $V/\sigma$ of all star-forming galaxies in TNG50 with $M_{\star} = (1-10)\times10^{10} M_\odot$ in our selected redshift range ($z$ = 3-5), as well as the recent data of $z\approx 4.5$ galaxies observed in the [CII] emission line using ALMA. 
The median $V/\sigma$ of the simulated galaxies is approximately 4.5 at $z = 3$ and gradually decreases with increasing redshift. 
The broadness of the 16th-84th percentile band exhibits a slight decrease with redshift. 
The blue crosses show galaxies with $V/\sigma$ more than 1-sigma above the median value, which we define as dynamically cold discs in the simulation. 
Several simulated galaxies exhibit relatively high $V/\sigma$ values, some exceeding 7, thereby overlapping the range of observed values \citep{lelli_massive_2021,rizzo2021dynamical}. We also present the velocity dispersion $\sigma$ distribution and compare it with that of observed galaxies in Appendix~\ref{appendix:velocity distrubution}.

\begin{figure}
\centering
\includegraphics[width=0.49\textwidth]{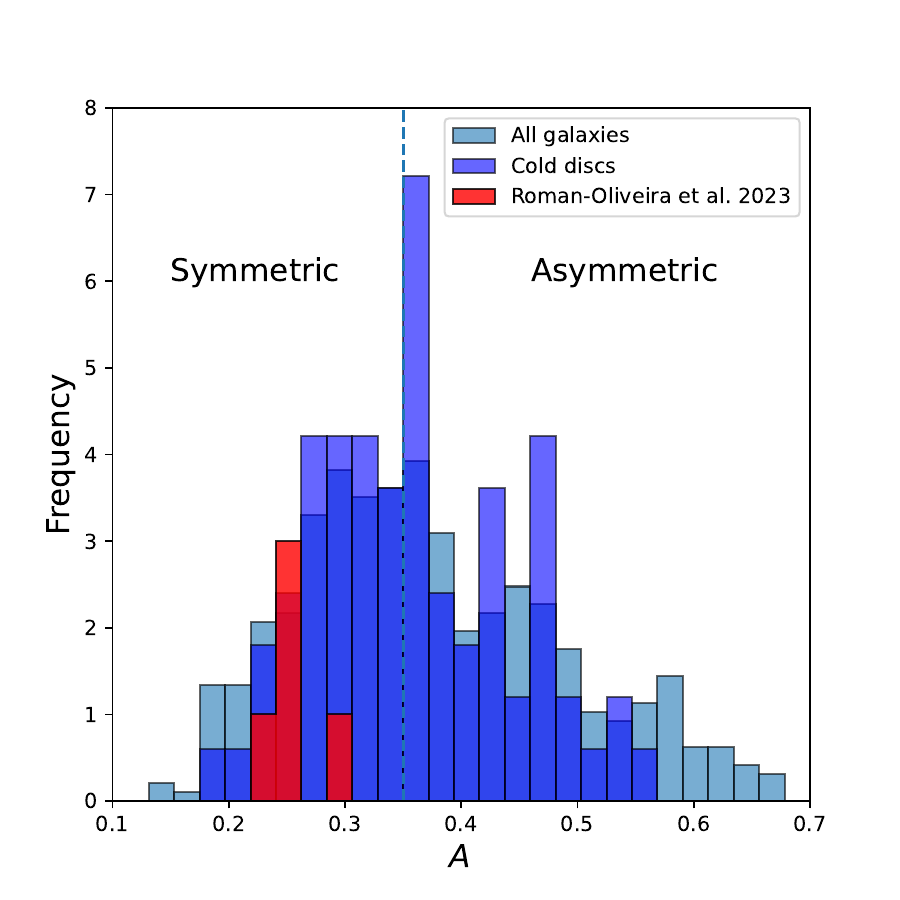}
\caption{Distribution of the asymmetry parameter $A$ for simulated cold disc galaxies, all star-forming simulated galaxies, and observed galaxies from \citet{2023MNRAS.521.1045R}. 
Galaxies with $A<0.35$ are classified as symmetric.}
\label{fig:asys}
\end{figure}

In Fig.~\ref{fig:asys}, we compare the distribution of the asymmetry parameter $A$ for the simulated galaxies to the values we have estimated for the real galaxies from \cite{2023MNRAS.521.1045R}. 
We show both the distribution of the cold discs (blue crosses in Fig.~\ref{fig:voversigma}) and that of all star-forming simulated galaxies in the selected mass range, and the two distributions are normalized as the area under the histogram sums to 1. 
For all simulated galaxies, the asymmetry parameter is widely distributed, from 0.1 to 0.7. 
The dynamically cold discs exhibit values ranging from 0.15 to 0.6, and a peak between 0.28 and 0.37, whereas the real galaxies in observation are concentrated between 0.2 and 0.32. 
We note that some simulated galaxies are highly symmetric despite not being classified as cold discs by the $V/\sigma$ ratio. 
We used $A = 0.35$ as an arbitrary threshold to classify galaxies into symmetric and asymmetric systems. 
The threshold is chosen on the basis of the $A$ values found in observed galaxies. 
Combining the $V/\sigma$ and $A$ classifications, we found that cold symmetric discs constitute about 10\% of our total simulated galaxy sample. 


\begin{figure*}
\begin{center}
\includegraphics[width=0.8\textwidth]{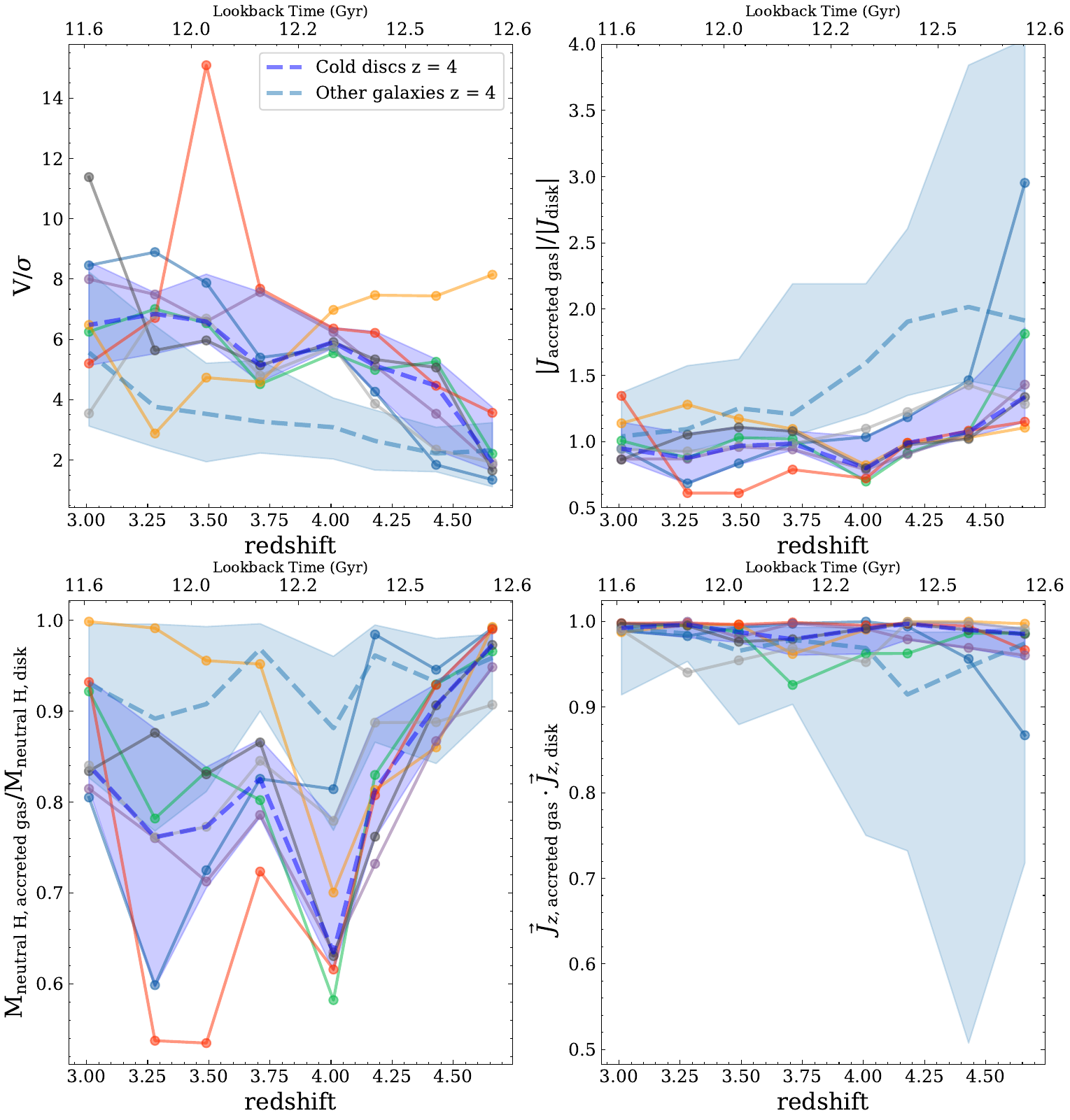}
\caption{Evolution of gas properties of simulated galaxies selected as cold discs at $z=4$ (coloured solid lines for single galaxies, deep blue dashed lines show the median value, deep blue shaded region for 16th to 84th percentiles) compared to the general TNG50 population of star-forming galaxies at $z=4$ (light blue dashed line for median, light blue shaded region for 16th to 84th percentiles). By gas, we always mean neutral hydrogen (see Section~\ref{subsection:Kinematic_definition}). Top Left Panel: $V/\sigma$ of the gas in the disc. Top Right Panel: ratio of the total angular momentum of the gas accreted between two subsequent snapshots over the angular momentum of the gaseous disc. Bottom Left Panel: mass ratio of new accreted gas over the disc gas at the current snapshot. Bottom Right Panel: misalignment between the angular momentum direction of accreted gas and the spin direction of the disc gas, measured as the dot product of two unit vectors (1 means perfect alignment).}
\label{fig:gas_accre}
\end{center}
\end{figure*}

\subsection{Evolution and assembly process of the TNG50 galaxies}
\label{subsection:evo of }

\subsubsection{Gas Properties}
Although galaxies with discs are a minority in the TNG50 simulation, their presence is notable because they were previously under-represented in earlier studies. This raises the important question: what differentiates these galaxies from the majority of others in the sample? Understanding this distinction is crucial for uncovering the conditions that lead to the formation of discs in the first place. To do this, we selected the TNG50 discs at $z=4$ (snapshot 21). Then, using the SubLink merger tree, we traced the progenitors and descendants. In Fig.~\ref{fig:gas_accre}, we plot the evolution of the gas properties of galaxies with discs at $z = 4$, along with other galaxies at $z = 4$ in our sample from $z = 4.75$ to $z = 3$. 
We omit the snapshot at $z = 5$ because our analysis focusses on gas accretion between subsequent snapshots. Thus, the earliest snapshot cannot be included due to the lack of a preceding time step. 
We chose this snapshot because most of the high-resolution ALMA [CII] data are in this range, and in TNG50 there are 7 cold disc galaxies and 39 star-forming galaxies in total at $z = 4$, which is a sample of sufficient size to perform a statistical study.

In the top left panel of Fig.~\ref{fig:gas_accre}, we show the evolution of $V/\sigma$ for the gas component of all galaxies selected at $z = 4$. 
Although most galaxies exhibit a general trend of increasing $V/\sigma$ as the redshift decreases, this pattern is not strictly monotonic for individual galaxies. 
On average, galaxies that are discs at $z=4$ have higher $V/\sigma$ values compared to other galaxies also before and after our selection, indicating that they remain dynamically cold over time. 
Starting from $z = 3.5$ to $z = 3$, the $V/\sigma$ values of discs display greater scatter and tend to join the distribution of other galaxies by $z = 3$. 
The galaxy with the larger $V/\sigma$ at $z=4$ (subhaloid = 12917) remains around a value 7 from $z = 4.75$ to $z = 4$, but it drops quickly and becomes comparable to other galaxies at $z = 3.71$. 
Another interesting example is subhaloid = 36491 at $z = 3.49$, which shows an unusually high $V/\sigma$ for just one snapshot, suggesting a transient phase lasting less than 200 Myr.

All galaxies in our sample are star-forming, which means they have a large gas reservoir and/or they experience copious gas accretion from the environment. 
As mentioned, the cold gas masses range from $10^{10}\ M_{\odot}$ to $10^{11}\ M_{\odot}$, with a slight tendency (factor 1.15 on average) for the dynamically cold discs to have more gas than normal galaxies.
Although these are large mass reservoirs, both theory \citep{Keres+2009,fraternali2012estimating} and the observational estimates of the depletion time \citep[e.g.][]{saintonge2011cold} indicate that gas accretion from the environment is crucial to feed star formation in galaxies at any redshift.
It is therefore important to investigate whether the properties of the accreted gas play a role in determining the dynamical state of the galaxies and the formation and/or maintenance of the discs. 
Here, we trace the gas from snapshot to snapshot by ParticleIDs, and the new accreted gas is defined as gas cells that appear inside the disc in snapshot $n$ but were absent in snapshot $n-1$. We define the height of the disc as one standard deviation of the density distribution along the z-axis, and the diameter of the disc as the galactocentric radius where the neutral hydrogen column density is 1 $M_{\odot} \rm pc^{-2}$. At $z = 4$, the mean heights of the TNG50 simulated discs are 0.63 pkpc and 0.88 pkpc for disc galaxies and all galaxies, respectively. 
The 1 $M_{\odot} \rm pc^{-2}$ neutral hydrogen radii are 8.52 pkpc and 9.56 pkpc for disc galaxies and all galaxies, respectively.
In the top right panel and the bottom two panels of Fig.~\ref{fig:gas_accre}, we illustrate the ratios of different properties between the accreted gas and the gas in the disc. 

The top right panel of Fig.~\ref{fig:gas_accre} shows the evolution of the ratio between the total angular momentum of the accreting gas over the total angular momentum of the galaxy disc after accretion. 
Clearly, the dynamically cold disc galaxies exhibit lower values compared to the other galaxies, particularly when the redshift is greater than $z=4.25$. 
This is largely due to the fact that cold discs accrete less gas than normal galaxies. 
This is visible in the bottom left panel of Fig.~\ref{fig:gas_accre}, which shows a noticeable difference in the mass ratio between newly accreted neutral hydrogen and the neutral hydrogen in the disc at the current snapshot, comparing discs and normal galaxies.
Finally, the bottom right panel displays the alignment between the specific angular momentum of the accreted gas and that of the galaxy disc after accretion. $\overrightarrow{J}_{z, \mathrm{accreted\ gas}} \cdot \overrightarrow{J}_{z, \mathrm{disc}}=1$ indicates perfect alignment and the subscript ‘$z$' denote the $z$-component of the angular momentum. 
This panel clearly highlights the differences in gas accretion pathways. 
For cold disc galaxies, the value is close to 1, which suggests that the accreted gas corotates with the galaxy upon reaching the galactic disc. This indicates that the alignment between the accreting gas and the galactic disc is a crucial factor in the formation and maintenance of disc galaxies in TNG50. We further complement this picture by showing the evolution of the scale height of newly accreted gas in Appendix~\ref{appendix:scaleheight}.

\begin{figure*}
\begin{center}
\includegraphics[width=0.8\textwidth]{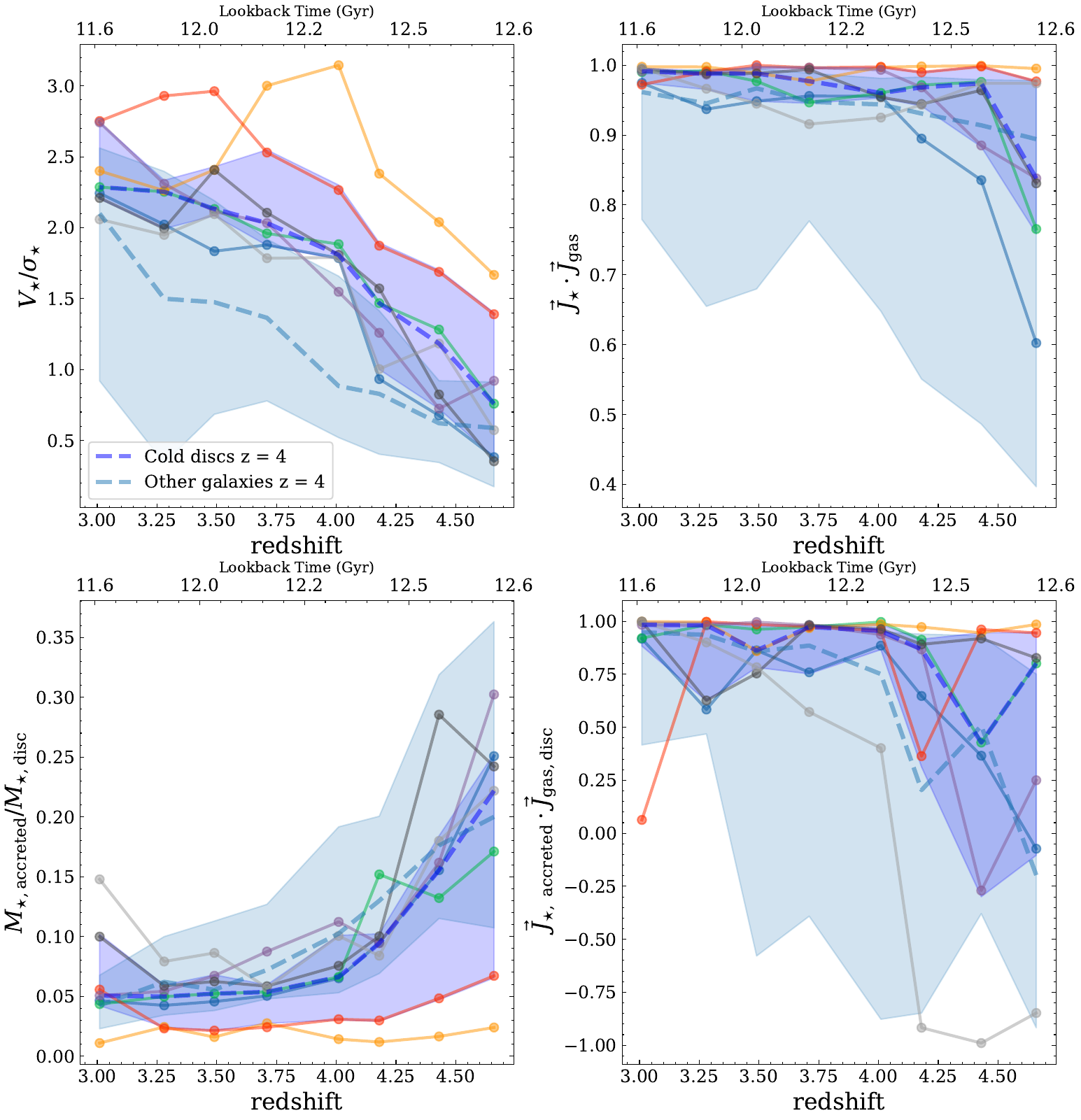}
\caption{Evolution of stellar properties of simulated galaxies selected as dynamically cold discs at $z=4$ (coloured solid lines for single galaxies, deep blue dashed lines show the median value, deep blue shaded region for 16th to 84th percentiles) compared to the general TNG50 population of star-forming galaxies at $z=4$ (light blue dashed line for median, light blue shaded region for 16th to 84th percentiles). Top Left Panel: $V/\sigma$ of the stellar disc over time. Top Right Panel: alignment between the spin direction of gas and stars in the disc, measured as the dot product of two unit vectors (1 means perfect alignment). Bottom Left Panel: ratio of accreted stellar mass over the mass of the stellar component in the disc. Bottom Right Panel: misalignment between the angular momentum direction of accreted stars and the spin direction of the galaxy stellar discs, measured as the dot product of two unit vectors.}
\label{fig:steller_evo}
\end{center}
\end{figure*}

\subsubsection{Stellar component}

From the above, we found that dynamically cold disc galaxies at $z\approx 4$ in TNG50 tend to accrete less gas than the average galaxy population. 
Moreover, this gas accretion in cold discs takes place with an orientation of the angular momentum aligned with the pre-existing gas disc.
Here, we investigate the properties of the stellar components of our simulated galaxies, including the properties of stars falling onto them due to merger events.
In Fig.~\ref{fig:steller_evo}, we present the evolution of the mass and angular momentum of the stellar component for cold disc galaxies selected at $z=4$, as well as other star-forming galaxies from the same epoch. We also show the $V/\sigma$ evolution of the stellar component. The markers used are the same as in Fig.~\ref{fig:gas_accre}. 
We select accreted stars by comparing the star-particle ids in two snapshots and exclude newly formed stars within that interval. 
The angular momentum of the stellar component is derived using all the star particles belonging to the galaxies. 

The top-left panel of Fig.~\ref{fig:steller_evo} illustrates the $V/\sigma$ ratio of the stellar disc over time. 
In the top-right panel, we show the dot product of the spin vectors of the stellar and gas components. 
The bottom-left panel displays the mass ratio between accreted stars and those in the disc, while the bottom-right panel depicts the dot product of the spin direction of the accreted stars and the stellar disc. 
Our findings show that galaxies hosting gas discs tend to exhibit higher stellar disc \( V/\sigma \) values, indicating more rotationally supported and kinematically colder stellar components. 
After \( z = 4 \), the stellar discs of these galaxies remain dynamically colder for a longer period compared to the gas disc. 
By \( z = 3 \), while the \( V/\sigma \) value of the gas disc of these galaxies converges toward those of the general population, their stellar discs still show a significant kinematic distinction. 
For most cold disc galaxies, the gas and stellar components tend to have an almost perfect alignment of their angular momentum. 
Moreover, this alignment persists in time from $z = 5$ to $z = 3$. 
These galaxies also tend to accrete fewer stars, with the accreted stars aligning with the disc plane, although this distinction is less prominent than for the gas component. 

\begin{figure*}
\begin{center}
\includegraphics[width=0.8\textwidth]{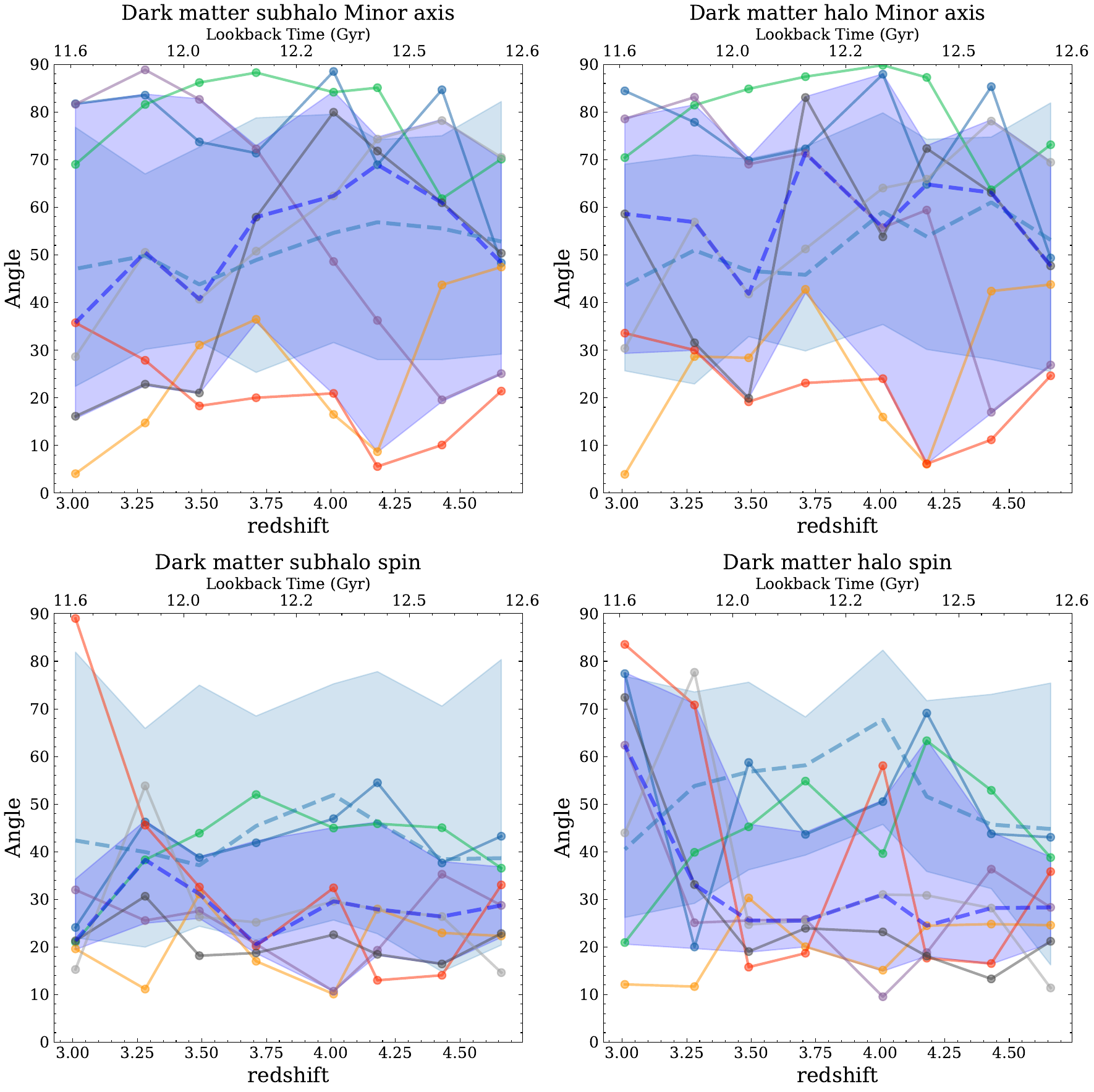}
\caption{Evolution of the angle between the spin direction of the gas in the galaxies and two reference axes of host dark matter subhalos and halos. 
Markers are the same as in Fig.~\ref{fig:gas_accre} for simulated galaxies selected as cold discs at $z=4$ compared to the general TNG50 population of star forming galaxies at $z=4$. Top Left Panel: angle between the spin direction of the galaxies and the minor axis of the dark matter subhalo. Top Right Panel: angle between the spin direction of the galaxies and the minor axis of the dark matter halo. Bottom Left Panel: angle between the spin direction of the galaxies and the spin direction of the dark matter subhalo. Bottom Right Panel: angle between the spin direction of the galaxies and the spin direction of the dark matter halo.}
\label{fig:dark_matter}
\end{center}
\end{figure*}

\subsubsection{Dark matter halo}
To explore the effects of the halo environment on the formation and maintenance of dynamically cold disc galaxies, we investigated the shape and angular momentum direction of the host dark matter halos and subhalos. 
Following \cite{2021MNRAS.504.6033S}, we use all the dark matter particles in the FOF halos and subhalos to characterise the shape of a dark matter halo by its mass tensor,
\begin{center}
\begin{equation}
I_{i j} \equiv \sum_{k=1}^N x_{k, i} x_{k, j},
\end{equation}
\end{center}
where $x_{k,i}$ represents the $i$-th component ($i$ = 1, 2, 3) of the position vector for the $k$-th dark matter particle, measured relative to the halo centre. The halo centre is defined by the position of the particle with the minimum gravitational potential energy. 
The shape and orientation are derived by the eigenvalue $\lambda_i\left(\lambda_1 \geqslant \lambda_2 \geqslant \lambda_3\right)$ and the corresponding eigenvector $\hat{e}_i$. 
The orientation of the minor axis of the halo is given by $\hat{e}_3$. 
The spin direction of the dark matter halo is calculated using all the dark matter particles in the halo. 
In Fig.~\ref{fig:dark_matter}, we present (top panels) the angles between the spin direction of the gas component and the major and minor axis of the host dark matter subhalos (left panels) and halos (right panels) at each redshift. 
For both the subhalos and halos, these average angles do not show significant differences between cold disc galaxies and other normal galaxies.
For some galaxies, the angles change rapidly over a short timescale. 
This could be caused by the small difference between the eigenvalue $\lambda_2$ and $\lambda_3$, which can cause the minor axis to transfer from one eigenvector to another eigenvector. 
The bottom panels of Fig.~\ref{fig:dark_matter} show that the spin alignment between gas and dark matter of cold disc galaxies is stronger than for the rest of the galaxy population, both for host halos and subhalos. 
The average angles for normal galaxies are 50$^\circ$, similar to the angle between the galaxy spin and the minor axis of the host halo. 
Whereas for cold disc galaxies, the angles are around 30$^\circ$ from $z = 5$ to $z = 3.5$, and increasing to be closer to the normal galaxies with the redshift decreasing. 
We conclude that the alignment between the spin of the galaxy and that of the hosting halo may play a positive role in the shaping and the maintenance of the discs in the TNG simulation \citep[e.g.][]{2020ApJ...895...17D,2022MNRAS.512.5978R,2023MNRAS.518.5253Y,2024ApJ...972...73S}.

\begin{figure}
\includegraphics[width=0.48\textwidth]{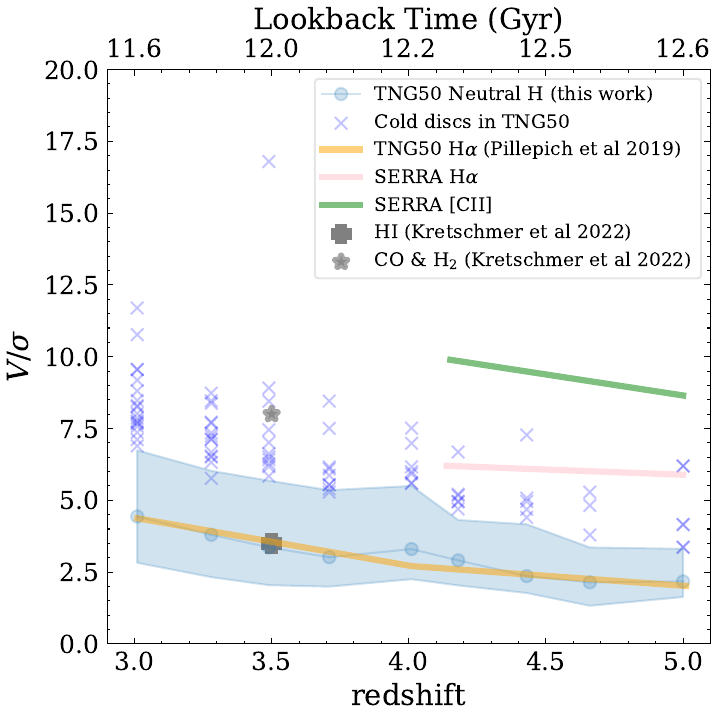}
\caption{Redshift evolution of $V/\sigma$ in this work, compared with results from other studies. The blue line and shaded region represent the median and 16th-84th percentiles of our sample. The blue crosses are cold discs. 
The yellow line corresponds to the median value for H$\alpha$ in the TNG50 simulations from \citet{Pillepich+2019}.
The pink and green lines represent H$\alpha$ and [CII] data from Serra simulations \citep{kohandel2024dynamically}, respectively. Gray plus signs and gray stars are the HI and CO $\&\ \mathrm{H}_2$ gas from a zoom-in simulation of a single galaxy in \citet{kretschmer2022origin}.
}
\label{fig:v/sigmacompare}
\end{figure}

\section{Discussion}
\label{section:discussion}

\subsection{Comparison to other works}
In \cite{Pillepich+2019}, a method different from ours is used to derive the $V/\sigma$ of the gas component in TNG50 galaxies. First, they focus on H$\alpha$ emitting gas and use the SFR to trace the H$\alpha$ emission. Instead, we select cold gas, specifically neutral hydrogen, as the tracer of the [CII] emission seen in ALMA observations. For the rotational velocity $V$ and the velocity dispersion $\sigma$, they calculate the absolute maximum of the rotation curve within twice the stellar half-mass radius, derived from an edge-on projection.
Although their method for estimating the velocity dispersion $\sigma$ is broadly similar to ours, they restrict it to gas cells located between one and two times the stellar half-mass radius, with a radial bin size of 0.5 ckpc. The rotational velocity $V$ is also derived within the same binning configuration. By contrast, we compute $V$ and $\sigma$ within a fixed radial aperture of 5 pkpc and adopt a radial binning of 0.7 pkpc, which closely matches the resolution of ALMA observational studies. The stellar mass range of their sample at $z= 3-5$ is between $10^{10}\ M_{\odot}$ and $10^{10.5}\ M_{\odot}$, similar to ours. Despite all these differences, as shown in Fig.~\ref{fig:v/sigmacompare}, their average $V/\sigma$ values align closely with ours.

Observationally, ionized gas, traced by H$\alpha$, shows higher velocity dispersion with respect to molecular and atomic gas \citep[e.g.][]{ubler_evolution_2019,girard2021systematic,rizzo2024alma}.
This is also seen in simulations when gas cooling below $10^{4}$ K is accounted for \citep{varidel2016resolved,girard2021systematic,2022MNRAS.514..480E,kohandel2024dynamically}. 
However, in TNG50 there is no distinction between gas components at temperatures below $10^{4}$ K.
Moreover, the SFR-based selection of star-forming gas used by \cite{Pillepich+2019} identifies both ionized gas and cold, dense molecular gas. 
Thus, it is likely that our selection based on temperature and density and their selection based on SFR returns essentially the same type of gas. 
Regarding the size of the discs employed in the work of \cite{Pillepich+2019}, it is important to notice that at high redshifts, the stellar components of galaxies are significantly more compact compared to the local universe. Compared with their work, our method measures $V/\sigma$ in a larger region. 

In \cite{kohandel2024dynamically}, $V/\sigma$ of galaxies with stellar masses above $10^{10} M_{\odot}$ is investigated from the Serra suite of zoom-in simulations. 
The circular velocity is used as the rotation velocity, which may slightly overestimate $V$, particularly for galaxies that lack a well-defined thin disc. 
\citet{kohandel2024dynamically} produced mock data of two emission lines: H$\alpha$ and [CII], the former as a tracer of warm ionised gas, and the latter of cold neutral atomic/molecular gas \citep[see][for details]{Grassi+2014, Kohandel+2020}. The line luminosities of H$\alpha$ and [CII] for each gas cell were obtained through post-processing with the spectral synthesis code CLOUDY \citep{2017RMxAA..53..385F}.
The ionised (H$\alpha$) and neutral ([CII]) components in their sample of simulated galaxies show average values of $V/\sigma$ of 6 and 9, respectively.
Given their choice to use the circular speed for $V$, the difference clearly arises from the different dispersions of the two gas components. 
Additionally, for [CII]-emitting gas, 60\% of their sample is found to be dynamically cold, with $V/\sigma$ values greater than 4. 
The $V/\sigma$ values from the Serra simulations are significantly higher than those of the neutral gas in TNG50. 
Although this discrepancy can be partially attributed to the choice of circular speed as rotation velocity, the Serra simulations differ from TNG50 in several aspects, such as resolution, numerical methods, and the physical treatment of star formation and feedback processes, which may also contribute higher $V/\sigma$ values.

\cite{kretschmer2022origin} identifies a dynamically cold disc galaxy in a zoom-in simulation performed using the adaptive mesh refinement code $\mathrm{ramses}$ \citep{2002A&A...385..337T}, where the formation of $\mathrm{H_2}$ and CO is modelled following the approximate chemical equilibrium prescription of \citet{2012MNRAS.421..116G}. The column densities of absorbing components such as dust, HI, and H$_2$ are approximated based on the local Jeans length of subgrid elements, assuming an isothermal condition at 10\,K. 
They found a measured $V/\sigma$ value for atomic HI gas of 3.5 at $z \sim 3.5$, consistent with the values found for the majority of star-forming galaxies in TNG50. In contrast, molecular gas tracers such as CO and H$_2$ reach $V/\sigma$ values as high as 8, comparable to the cold disc values we found in TNG50 and observational estimates for high-redshift galaxies observed with ALMA. 
Similarly to \cite{kohandel2024dynamically}, the differences between HI and the molecular gas in \cite{kretschmer2022origin} underscore the varying kinematic properties within the multiphase nature of the interstellar medium. 

In this work, we also investigated the lifespan of cold disc galaxies and their evolution over time in TNG50. Our findings show that, for the galaxy with peak $V/\sigma$ values exceeding 10 ($\approx3$ sigma above the median, see Fig.~\ref{fig:gas_accre} top left panel), the cold disc phase is relatively short-lived, with an upper limit of around 200 Myr, constrained by the time interval between consecutive simulation snapshots. However, for values of $V/\sigma$ between 1 and 2 sigma above the median, which constitutes the majority of the cold disc sample, they consistently exhibit such $V/\sigma$ values above the average in four snapshots, covering redshifts $z=3.25-4.4$, corresponding to a time of approximately 600 Myr. 
This indicates that, while the phase of extreme rotational support is short, for the larger population of galaxies with moderately cold discs ($V/\sigma$ between 5 and 10), which are similar to observed $z\approx4.5$ galaxies, the timescales are much longer.

The short-lived nature of the most extreme cold discs is consistent with previous findings. For example, \citet{kretschmer2022origin} reported a disc phase lasting around 400 Myr in their zoom-in simulations. However, in that case, the "cold" state with a $V/\sigma$ value comparable to our threshold (i.e., $V/\sigma \gtrsim 5$) appears only in a single snapshot. 
Although individual cases like this have been presented, the long lifespan of moderately cold discs was not clearly identified before. In TNG50, we find that this prolonged lifespan is closely linked to the accretion histories of the galaxies, as we discuss in more detail in Section~\ref{subsection:discform}.


\subsection{How do cold disc galaxies form in TNG50?}
\label{subsection:discform}
Until recently, the general theoretical expectation was that dynamically cold discs are not the predominant galaxy structure at $z>2$. 
Most studies \citep[e.g.][]{dekel2014wet, 2015MNRAS.450.2327Z, 2017MNRAS.465.1682H, Pillepich+2019} suggest that high-redshift galaxies typically exhibit significant kinematic turbulence, hindering the formation of stable cold discs.
These findings have been challenged by ALMA observations of numerous regularly rotating gas discs now seen up to $z=7$ \citep{neeleman_cold_2020, lelli_massive_2021,rizzo_dynamically_2020,fraternali_fast_2021,rizzo2021dynamical,2023MNRAS.521.1045R,rowland2024rebels} and, more recently, by JWST which has found a high fraction of stellar discs at those redshifts \citep[e.g.][]{kartaltepe2023ceers}. 
Therefore, it is very relevant that we found the presence of discs in TNG50. 
Despite being rare, this indicates that current cosmological hydrodynamical simulations can form disc galaxies at $z$ > 2 with properties very similar to the observed ones. We can then investigate how different they are with respect to the common population of chaotic galaxies in TNG50 with the aim of better understanding their formation (and maintenance) processes.

\begin{figure*}
\includegraphics[width=\textwidth]{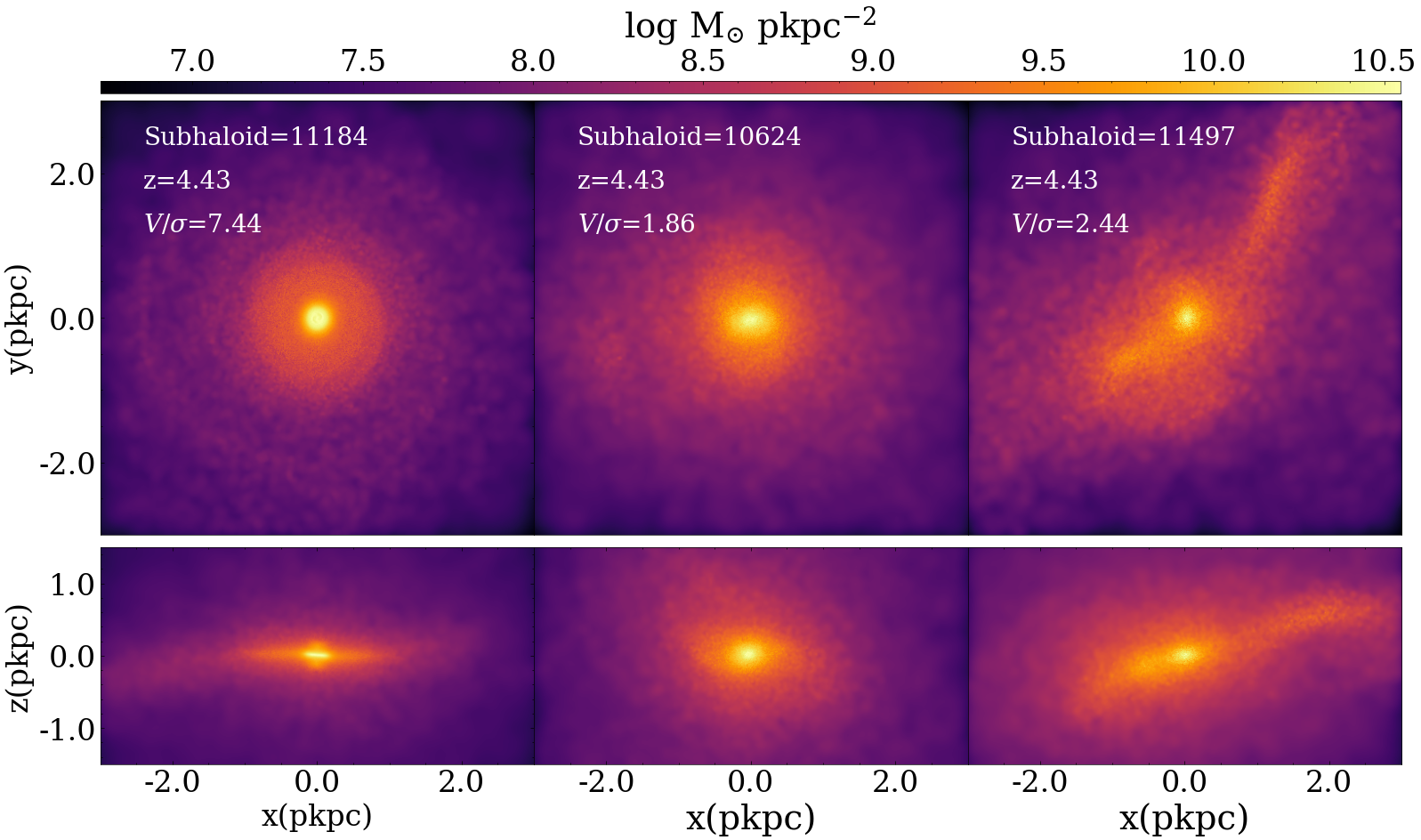}
\caption{Stellar component in three simulated galaxies presented in Fig.~\ref{fig:fig1}. 
The figure is also rendered by Py-SPHViewer code \citep{benitez-llambay_2015}.}
\label{fig:stardensity}
\end{figure*}

As discussed in Section~\ref{subsection:evo of }, one key mechanism for the formation of a rotation-dominated cold gas disc is the recent co-planar and co-rotating accretion of gas \citep[see also][]{scannapieco2009formation,sales2012origin,hafen2022hot}. 
Our findings also suggest that dynamically cold disc galaxies tend to accrete less gas compared to other galaxies, reducing the impact of misaligned gas clumps that could disrupt rotational coherence. 
The visual inspection of the gas components in a large number of simulated galaxies indicates that gas accretion occurs via large and massive complexes, clouds or filaments, and these complexes seem responsible for the disruption of the discs, in particular when they fall in with misaligned spin with respect to the central galaxy.
An interesting possibility is that the presence of such large gas complexes is due to the low mass resolution of TNG50 in the circum-galactic medium, something that could be alleviated and improved in future simulations \citep{2020MNRAS.498.2391N,2023MNRAS.522.1535R}.
A similar conclusion was also reached when comparing TNG50 with local spiral galaxies \citep{Marasco+2025}.

Furthermore, high fractions of star formation in situ support the development of cold discs, as stars form directly from accreted gas, thus helping to preserve the rotational structure \citep{scannapieco2011formation}. 
Conversely, a high merger rate due to infalling companion galaxies can lead to the disruption of a disc in formation.
In this context, we examined the stellar component in our example galaxies shown in Fig.~\ref{fig:fig1}, finding that their stellar morphology closely resembles that of their gas component.
We show this in Fig.~\ref{fig:stardensity}, where we observe that the stellar component of the cold disc galaxy forms an extended, thin disc, resembling the structure of local spiral galaxies. 
There are subtle indications of spiral arms, along with a symmetric warp in the outer disc. 
In addition, there appears to be a small central bulge. 

The stellar components of the other two galaxies, representative of the normal TNG50 population, exhibit markedly different structures. 
In the middle panel, the galaxy displays a compact structure with a smooth envelope, indicative of a small and dynamically hotter stellar disc. 
There are no visible signs of mergers; instead, the system appears puffy and ellipsoidal, probably due to misaligned gas accretion, which can inflate the stellar distribution and suppress the formation of an extended, ordered disc. This galaxy exemplifies the common case in TNG50 where chaotic gas accretion, rather than mergers, drives kinematic irregularities in both the gas and stellar components. 

Finally, the right panel of Fig.~\ref{fig:stardensity} shows clear evidence of ongoing interactions and possibly a merger event. Although a compact central stellar component is still visible, the presence of two distinct substructures surrounding the galaxy highlights typical signatures of galaxy interactions. Such interactions disturb the stellar disc, creating irregular features and disrupting the rotational support. Although mergers are less frequent than chaotic accretion in this mass range, they remain an important process in shaping galaxy dynamics, as also seen in observational samples \citep[e.g.][]{2023MNRAS.521.1045R}. 
Notably, observations of high-redshift galaxies with ALMA \citep{rizzo_dynamically_2020,neeleman_cold_2020,lelli_massive_2021,fraternali_fast_2021,rizzo2021dynamical,2023MNRAS.521.1045R} tend to show gas kinematic patterns typical of the stellar cold disc case shown here (left), reinforcing the importance of further investigating the conditions required for the formation and long-term survival of dynamically cold discs.

\begin{figure}
\includegraphics[width=0.45\textwidth]{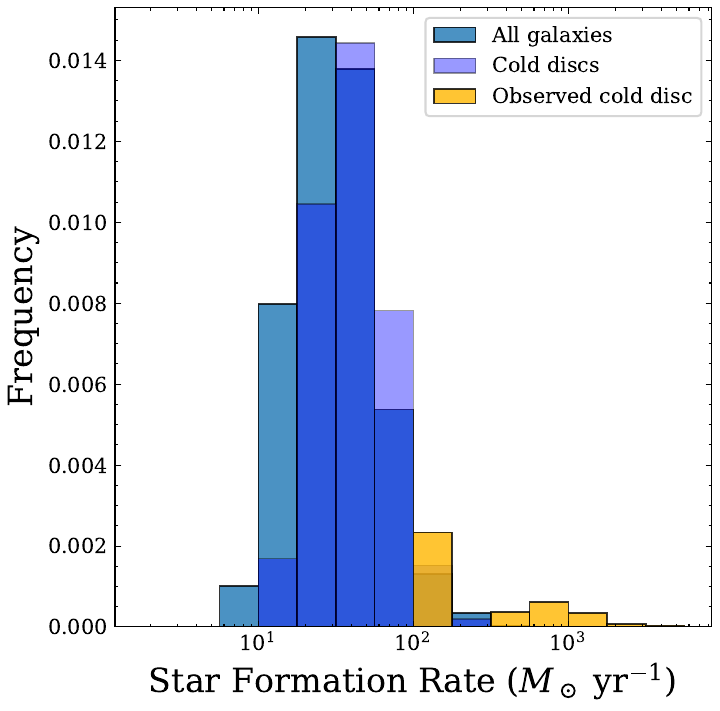}
\caption{Normalized star formation rate distribution of all TNG galaxies in our samples, galaxies with cold discs and observed galaxies from \citealp[]{rizzo_dynamically_2020,rizzo2021dynamical,2023MNRAS.521.1045R}.}
\label{fig:starformingrate}
\end{figure}

Given that some observed disc galaxies are classified as starbursts, we also examined the SFR of the dynamically cold discs in our sample, shown in Fig.~\ref{fig:starformingrate}. Compared with the whole star-forming galaxies samples, the distributions exhibit a similar overall trend. Notably, we do find a few cold disc galaxies with a star formation rate exceeding 150 $M_{\odot}$ yr$^{-1}$ but less than 300 $M_{\odot}$ yr$^{-1}$, indicating that high star formation activity is not entirely absent in cold disc galaxies. 
However, the highest SFR of the whole samples ($\approx 400$ $M_{\odot}$ yr$^{-1}$) is significantly lower than that of the most actively star-forming galaxies observed in high-redshift observations \citep[e.g.][]{2017ApJ...849...45C,2023MNRAS.521.1045R}. 
This could be partially attributed to the highly efficient ejective feedback in TNG50, which removes gas more rapidly than it occurs in reality, thus inhibiting its accumulation and a high SFR\citep[e.g.][]{2019MNRAS.490.3234N,2021MNRAS.502.2922H}. 
However, other biases may also play a role, for instance the time sampling of the snapshots may miss short moments of intense star formation activity. 
The lack of very high SFRs in TNG simulations has been pointed out in other works \citep{Lim+2021,Shen+2022} and a detailed investigation of the reasons behind it is beyond the scope of this paper.

It is worth noting that our main selection criterion, based on stellar mass, is unaffected by the lower SFRs in simulations compared to observations. 
In practice, this criterion naturally selects all TNG50 galaxies with the highest SFRs within the chosen redshift range. 
Since \citet{rizzo2024alma} found that cold gas velocity dispersion correlates with SFR, the systematically lower SFRs in simulated galaxies should bias us toward lower dispersions relative to observations. 
Instead, we find the opposite: TNG50 galaxies show significantly higher velocity dispersions on average, reinforcing our main conclusion.

\subsection{The future of high-redshift cold disc galaxies}

Given the large stellar masses (>$10^{10}\ M_{\odot}$) and the high SFR, the high-$z$ galaxies that we are investigating in this work are expected to evolve into massive present-day early-type galaxies (ETGs), whose stellar kinematics is dominated by random motion, i.e.\ slow rotators \citep[e.g.][]{2011MNRAS.414..888E,2014ApJ...782...68T,2016ARA&A..54..597C}. 
However, the extreme high rotation velocities observed in some $z\sim4$ galaxies pose the question whether they could evolve into massive disc galaxies, called super-spirals \citep{2018A&A...612L...6P,ogle2019break} or into local fast rotators \citep{2011MNRAS.414..888E,2016ARA&A..54..597C}. 
In \cite{rizzo_dynamically_2020}, they demonstrated that if the entire gas reservoir of their $z=4.2$ galaxy were converted into stars, the resulting galaxy would closely resemble fast rotators in the local universe in terms of size-stellar mass, circular speed-stellar mass, dark matter fraction-stellar mass relations. 
However, the most massive $z\sim4$ galaxies clearly overlap with the most massive local ETGs (slow rotators) in the circular speed-baryonic mass plane, indicating that these high-$z$ galaxies are nearly fully formed ETGs, except for their rapid rotation, which requires an important transformation \citep{fraternali_fast_2021, Roman-Oliveira+2024}. 
\cite{rizzo2021dynamical} investigated the size-stellar mass relation for their $z\sim4$ galaxies compared to local ETG and compact star-forming galaxies (cSFG). 
They found one galaxy with a relatively low gas content that already lies in the local ETG size-stellar mass relation. If the entire gas reservoir were converted into stars, the stellar mass of $z\sim4$ galaxies in their sample would reach the typical mass range for local ETGs or cSFGs.

In this study, we have established that TNG50 is capable of producing galaxies that closely resemble the ALMA galaxies observed at $z\approx 4$. 
This provides a unique opportunity to follow the future evolution of these galaxies down to $z = 0$ and investigate whether they are the progenitors of ETGs or other galaxy types. The total number of cold-disc descendants in our sample is 22. 
This number is relatively small, because in some galaxies, the gas remains in the cold disc phase for a long period of time. As a result, the same galaxy can be selected as a cold disc in multiple snapshots, and when traced to $z=0$, these duplicate selections converge to the same descendant galaxy.
The final stellar masses of all cold disc descendants range at $z = 0$ from $8 \times 10^{10}\,M_{\odot}$ to $5 \times 10^{12}\,M_{\odot}$. 

By tracing the evolution of our galaxies to $z=0$, we match the galaxies with the TNG disc Galaxy Kinematic Decompositions catalogues (\citealp{2019ApJ...884..129D,2020ApJ...895..139D}), which include disc galaxies with stellar masses greater than $10^{10} M_{\odot}$ and more than half of their kinetic energy in ordered rotation. We find that 36\% of our sample appear in the catalogue, which means that one third of the high-redshift cold disc galaxies in TNG50 are the progenitors of massive disc galaxies.
These are galaxies with typical SFRs of $0-{\rm a\,few}\,M_{\odot}\,{\rm yr}^{-1}$.
At $z=0$, these massive disc galaxies occupy the lower mass regime of the cold disc descendants (i.e., $M_{\star} < 3 \times 10^{11} M_{\odot})$. 
This suggests that they have not experienced significant external accretion and merger events and have retained their rotational support. 
On the other hand, descendants with higher stellar masses appear to have accreted large amounts of gas or suffered more frequent merger events.

Following the fast/slow-rotator classification of \citet{2016ARA&A..54..597C}, which combines the $\lambda_{Re}$ parameter (the stellar spin proxy measured within the effective radius; \citealp{2007MNRAS.379..401E,2011MNRAS.414..888E}) with the ellipticity of the galaxy, we find only one slow rotator among our descendants of high-$z$ dynamically cold discs.
The low fraction of slow rotators suggests that most descendants of cold discs maintain their rotation-dominated structures despite external perturbations. 

\section{Conclusions}
\label{section:conclusions}
In this paper, we have studied the occurrence of galaxies with dynamically cold discs at high redshifts in the cosmological hydrodynamical simulation TNG50.
We selected star-forming galaxies in the redshift range $3<z<5$ and with stellar masses $M_{\star}>10^{10}\,M_{\odot}$ to ensure a fair comparison with galaxies observed with ALMA in the [CII] emission line at similar redshifts. 
We built mock emission-line datacubes from the simulated galaxies and analysed them in the same way as was done for real observations.
Furthermore, we investigated the formation and evolution of cold disc galaxies in TNG50 by examining the kinematics of gas, stars, and dark matter components. 
The main results of this study are summarised as follows:
\begin{enumerate}
\item
We found some star-forming massive galaxies with cold discs in TNG50, characterised by $V/\sigma > 5$, which is comparable to the values found in observations.
The majority of TNG50 massive star-forming galaxies at $z\sim 4$ are, however, chaotic systems, with $V/\sigma$ values around $2-3$. 
\item
By creating mock emission-line datacubes for the neutral hydrogen component in simulated galaxies and applying observational methods ($\mathrm{^{3D}Barolo}$) to derive $V/\sigma$, we confirmed that the values obtained closely match those directly derived from the simulation. 
This consistency validates the ability of codes like $\mathrm{^{3D}Barolo}$ to accurately recover the kinematic information of disc galaxies at these redshifts. 
\item
Our mock datacubes allow us to compare the cold gas kinematics of simulated discs with observed galaxies in detail. 
Out of this comparison, we found a good resemblance, in particular in the shape, extent and kinematic pattern of the position-velocity diagrams. 
This indicates that disc galaxies with properties very similar to the observed ones can form in TNG50.
\item 
We used mock datacubes to calculate a 3D asymmetry parameter, $A$, and compare it with the values obtained from real observations. Although observed galaxies exhibit greater symmetry on average, we found a subset of simulated galaxies with asymmetry parameters similar to those of their observed counterparts.
This parameter should be used in combination with $V/\sigma$ to characterise the regularity of a galaxy both morphologically and kinematically. 
Overall, we found that about 10\% of the simulated galaxies have $A$ and $V/\sigma$ in line with current observations. 
\item
We investigated the time evolution of simulated discs in TGN50 compared with the general population of simulated galaxies. 
 Disc galaxies with $V/\sigma$ between 5 and 10 remain dynamically colder than the general population of galaxies for most of the time from $z = 5$ to $z = 3$. 
 However, the extremely high $V/\sigma$ values (above 10) persist only for $\sim$ 200 Myr. 
\item
We find that disc galaxies accrete lower gas mass compared to the general galaxy population. Moreover, the new accreted gas is predominantly co-planar and co-rotating with the pre-existing disc. This combination of reduced and aligned gas accretion is a key factor in maintaining the stable disc structure. We speculate that the massive accreting gas complexes, hampering the formation of stable discs, can be due, at least partially, to the low resolution that TNG50 achieves in the circumgalactic medium. 
\item
Galaxies with dynamically cold gas discs also have colder stellar discs, which persist until at least $z = 3$. Limited merging with other stellar systems along with alignment of the stellar and gaseous discs are also crucial for both the formation and maintenance of the gas disc.
Additionally, the spin direction of the dark matter subhalos of disc galaxies with high $V/\sigma$ shows greater alignment with the disc orientation with respect to the average population.
\item
We traced all the dynamically cold disc galaxies at $z\sim 4$ in TNG50 down to $z = 0$ and found that 36\% are the progenitors of massive disc galaxies, while the others become fast rotators and one galaxy ($\sim5\%$) transitions into a present-day slow rotator.

\end{enumerate}

In conclusion, we reported the existence of disc galaxies with $V/\sigma>5$ at $z\approx 4$ in the cosmological hydrodynamical TNG50 simulation and with gas properties very similar to those observed by ALMA in the [CII] emission line.
Such discs are, however, the minority of the simulated galaxies, as the general population shows chaotic kinematics and large-scale morphological asymmetries.
Despite a precise statistic of the fraction of regularly rotating discs currently lacking in observations, the majority of observed galaxies with sufficient resolution in the mass range above $M_{\star}\sim 10^{10} \, M_{\odot}$ are rotating discs in their gas component.
Thus, the paucity of these systems in TNG50 should be understood. 
Our investigation indicates that it is the copious accretion of misaligned gas and occasional mergers that hamper the formation of stable discs and disrupt existing ones in TNG50 at $z\approx 4$. 
In the rare cases in which these events do not take place, a disc can remain in place for several hundreds of Megayears.

\section*{Acknowledgements}
\addcontentsline{toc}{section}{Acknowledgements}
We thank an anonymous referee for a constructive report, which has improved the quality of our paper.
We acknowledge the helpful discussions with Kyle Oman and Annalisa Pillepich. 
This work has received funding from the National Natural Science Foundation of China (NSFC No. 12425303) and the European Research Council (ERC) under the Horizon Europe research and innovation programme (Acronym: FLOWS, Grant number: 101096087). 

\section*{Data Availability}
The IllustrisTNG simulations, including TNG50, are publicly available and accessible at www.tng-project.org/data. Data directly referring to content and figures of this publication are available upon reasonable request from the corresponding author.



\bibliographystyle{mnras}
\bibliography{ref} 




\appendix

\section{$\mathrm{^{3D}Barolo}$ performance in retrieving the correct $V$ and $\sigma$}
\label{appendix:apenA}

\begin{figure*}
\centering
    \includegraphics[width=0.32\textwidth]{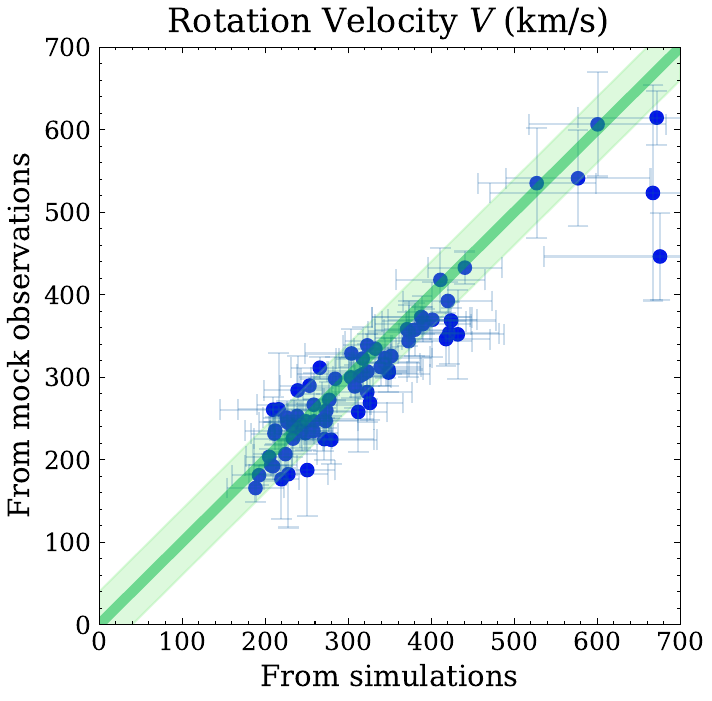}
    \includegraphics[width=0.32\textwidth]{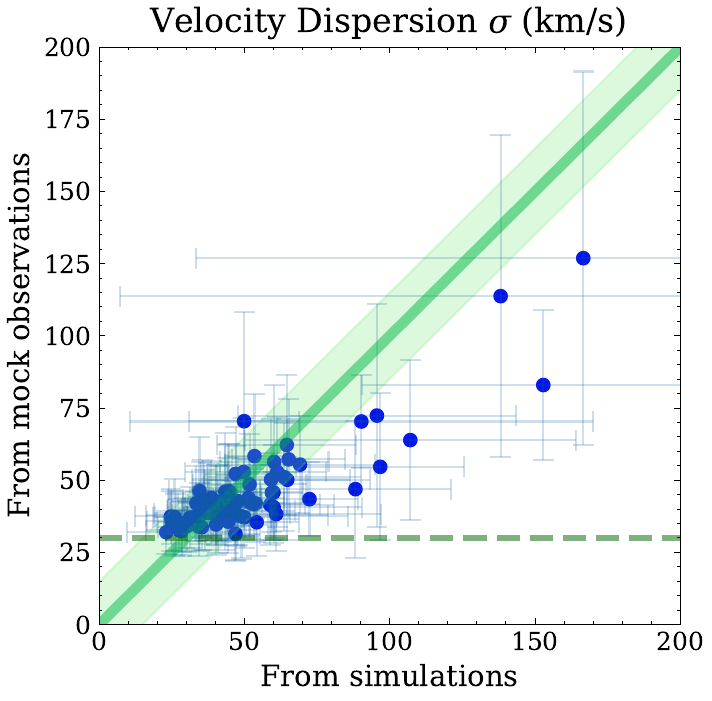}
    \includegraphics[width=0.32\textwidth]{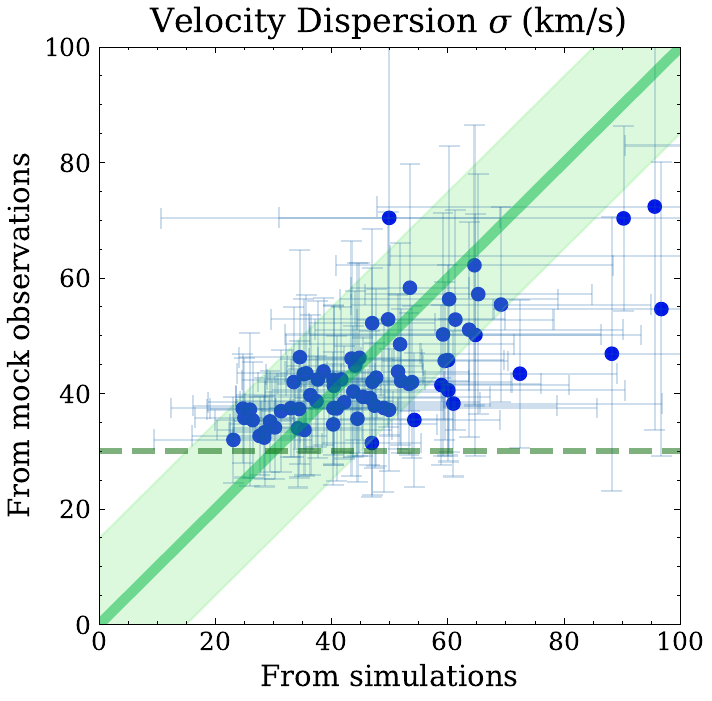}
\caption{Comparison of $V$ (left) and $\sigma$ (middle and right, in different range) values of simulated galaxies with dynamically cold discs, obtained either directly from the particles (x-axis) or from the mock datacubes (y-axis). Symbols are same as Fig.~\ref{fig:voversigma_mockvssim}. The green dashed lines show the spectral resolution of our datacubes.}
\label{fig:threeinrow}
\end{figure*}

Given the potential degeneracy in $V/\sigma$, we assess the recovery of $V$ and $\sigma$ separately in Fig.~\ref{fig:threeinrow} to evaluate the accuracy of the $\mathrm{^{3D}Barolo}$ modeling for cold disc galaxies. The left panel shows good agreement in rotation velocity $V$ between the mock datacubes and the simulations. For the velocity dispersion $\sigma$, the middle and right panels present results over two different ranges: the middle panel includes all galaxies ($\sigma$ = 0–200 km/s), while the right focuses on the majority with $\sigma$ = 0–100 km/s.
Compared to the rotation velocity, the dispersion values show larger scatter. This is mainly due to the limited spatial resolution of the mock datacubes and the fact that rotation dominates the kinematics in these systems.

Most galaxies follow the one-to-one relation, but several outliers are present. The main reason for these deviations is that these galaxies are very small in size, with only two spatial resolution elements across the disc, making it difficult to recover accurate dispersion values, which is reflected in their large error bars. Additionally, we did not mask the central region. The presence of central black holes can significantly enhance the velocity dispersion in the inner parts, leading to further discrepancies.
To assess the impact of the central region, we performed a test excluding the innermost data points. This resulted in a modest reduction in the measured dispersion and a corresponding increase in $V/\sigma$ for the cold disc galaxies. However, this change has little effect on the overall trends for the majority of high-redshift star-forming galaxies.
Finally, in our fit we do not allow $\sigma$ to have values lower than the spectral resolution of the datacubes, dashed horizontal line in the middle and right panels of Fig.~\ref{fig:threeinrow}.

\section{Velocity dispersion distribution of simulated galaxies}
\label{appendix:velocity distrubution}

\begin{figure}
\centering
    
    \includegraphics[width=0.5\textwidth]{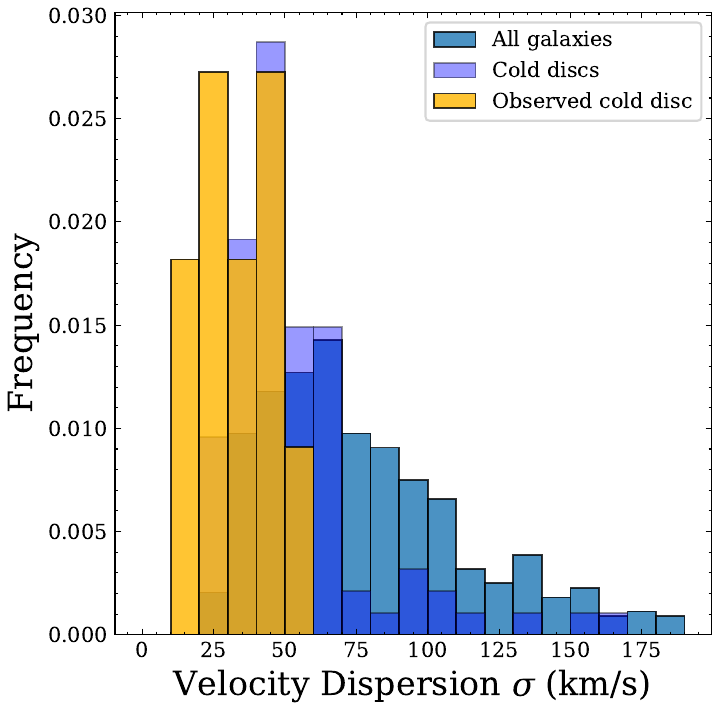}
\caption{Normalized velocity dispersion distribution of all TNG50 galaxies in our samples, galaxies with cold discs and observed galaxies from \citet{lelli_massive_2021}, \citet{rizzo2021dynamical}, and \citet{2023MNRAS.521.1045R}.}
\label{fig:VelocityDispersionob}
\end{figure}

Here we compare the velocity dispersion ($\sigma$) distribution of the simulated galaxies with observational measurements. 
Figure~\ref{fig:VelocityDispersionob} shows the normalized velocity dispersion distributions for all TNG50 galaxies in our sample, the subset classified as cold discs, and observational measurements from \citet{lelli_massive_2021}, \citet{rizzo2021dynamical}, and \citet{2023MNRAS.521.1045R}. 
The velocity dispersions of the TNG50 galaxies span a wide range, with the full sample extending to significantly higher values than those typically observed. 
The subset of galaxies with cold discs shows lower dispersions compared to the full simulated sample, but the observed galaxies still exhibit a slightly larger fraction of systems with low velocity dispersion.

\section{Scale height of new accreted gas}
\label{appendix:scaleheight}

\begin{figure}
\centering
    
    \includegraphics[width=0.5\textwidth]{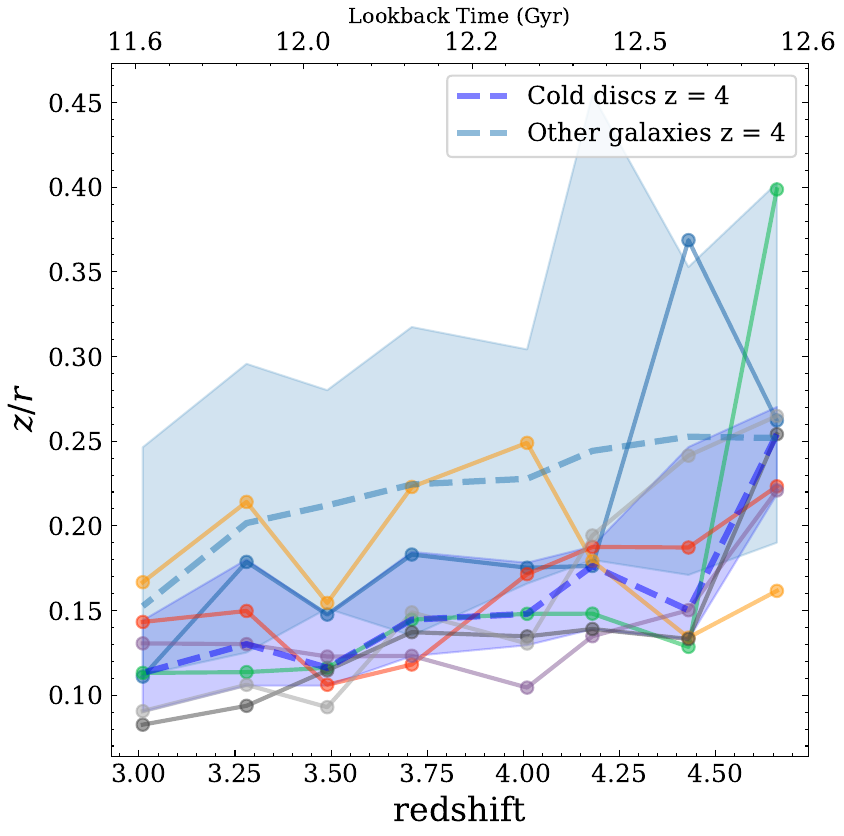}
\caption{Evolution of the scale height of newly accreted gas in simulated galaxies. The galaxy sample is the same as in Fig.~\ref{fig:gas_accre}. Cold disc galaxies are shown as dark coloured lines for individual galaxies, with the dark blue dashed line indicating the mean value and the dark blue shaded region representing the 16th to 84th percentile range. For comparison, the general TNG50 population of star-forming galaxies at $z = 4$ is shown as a light blue dashed line (median) with a light blue shaded region (16th to 84th percentiles).}
\label{fig:scaleheight}
\end{figure}

In Fig.~\ref{fig:scaleheight}, we present the evolution of the scale height of newly accreted gas for all galaxies selected at $z = 4$, following the same definition of newly accreted gas as in Fig.~\ref{fig:gas_accre}. The scale height is computed as the vertical height of each newly accreted gas particle divided by its projected distance to the galaxy center on the disc plane. We calculate the average scale height weighted by the neutral hydrogen mass of the gas particles. Most cold disc galaxies accrete gas with a low scale height, indicating that the accretion is well aligned with the disc plane.

\bsp	
\label{lastpage}
\end{document}